%% file: paper.tex
\documentclass[a4paper,
]{scrartcl}
\usepackage{url}
\newcommand{\email}[1]{\url{#1}}
\usepackage[utf8x]{inputenc}
\usepackage[T1]{fontenc}
\usepackage{xcolor}
\usepackage{multicol}
\usepackage{listings}
\usepackage{amsfonts} 
\usepackage{amssymb}
\usepackage{amsmath}
\usepackage{amsthm}
\usepackage{mathtools}
\usepackage{braket}
\usepackage{textcomp}
\usepackage{pifont}
\usepackage{bbding}
\usepackage{stmaryrd}
\usepackage{graphicx}
\usepackage[english]{babel}
\usepackage{proof}
\usepackage{tikz}
\usetikzlibrary{arrows,automata,positioning}
\tikzset{
  state/.style={
    rectangle,
    rounded corners,
    draw=black,  thick,
    minimum height=2em,
    inner sep=1pt,
    text centered,
  }
}

\usepackage[
 plainpages=false,
 pdftitle={A Complexity Preserving Transformation from Jinja Bytecode to Rewrite
Systems},
 pdfauthor={Georg Moser and Michael Schaper},
]{hyperref}
\usepackage{jbc}

\newtheorem{theorem}{Theorem}[section]
\newtheorem{corollary}{Corollary}[section]
\newtheorem{lemma}{Lemma}[section] 

\newtheorem*{claim*}{Claim}

\theoremstyle{definition}      
\newtheorem{definition}{Definition}[section]

\newtheorem{example}{Example}[section]

\deffootnote{1em}{1em}{\textsuperscript{\thefootnotemark\ }}

\title{A Complexity Preserving Transformation from Jinja Bytecode to Rewrite
Systems}
\author{%
Georg Moser\\
Institute of Computer Science,\\
University of Innsbruck, Austria\\
\email{georg.moser@uibk.ac.at} \and
Michael Schaper\\
Institute of Computer Science,\\
University of Innsbruck, Austria\\
\email{michael.schaper@uibk.ac.at}
}

\begin{document}

\maketitle

\begin{abstract} 
We revisit known transformations from Jinja bytecode
to rewrite systems from the viewpoint of runtime complexity. Suitably
generalising the constructions proposed in the literature, we define an
alternative representation of Jinja bytecode (JBC) executions as
\emph{computation graphs} from which we obtain a novel 
representation of JBC executions as \emph{constrained rewrite systems}. 
We prove non-termination and complexity preservation of 
the transformation. We restrict to
well-formed JBC programs that only makes use of non-recursive methods.
Our approach allows for simplified correctness proofs and provides a framework 
for the combination of the computation graph method with standard
techniques from static program analysis.
\end{abstract}

\input{introduction}

\input{preliminaries}
\input{jinja}

\input{abstraction}

\input{computationgraph}

\input{rewriting}

\input{implementation}

\input{conclusion}

\bibliographystyle{abbrv} 

\newpage
\appendix

\input{appendix}

\end{document}

%% file: introduction.tex
\section{Introduction}

In recent years research on complexity of rewrite systems has matured
and a number of noteworthy results could be established. We give
a quantitative assessment based on the annual competition of 
complexity analysers within TERMCOMP.%
\footnote{\url{http://termcomp.uibk.ac.at/}.}
With respect to last year's run of TERMCOMP, we see a success rate of
38 \% in the category \emph{Runtime Complexity – Innermost Rewriting}.
Note that the corresponding testbed is not restricted to polynomial runtime complexity
in any way. With respect to a qualitative assessment we want
to mention the very recent efforts to apply methods from linear
algebra and automata theory to complexity~\cite{MMNWZ11}; recent
efforts on adaption of the dependency pair method to complexity~\cite{HM08,HM08b,NEG11,HM11}
and the ongoing quest to incorporate compositionality~\cite{ZK10,AM13}.
(See~\cite{M09} for an overview in methods of complexity analysis of term rewrite systems.)

In this paper we are concerned with the applicability of these
results to automated runtime complexity analysis of
imperative programs, in particular of Jinja bytecode (JBC) programs.
Jinja is a Java-like language that exhibits the core
features of Java~\cite{SSB2001}. Its semantics is clearly defined and machine
checked in the theorem prover Isabelle/HOL~\cite{KN06}.

We establish a \emph{complexity preserving} transformation from
JBC programs $\Program$ to \emph{constrained term rewrite systems}
$\RS$, that is, the runtime complexity function with respect to
$\Program$ is bounded by the runtime complexity function with respect to
$\RS$ (Theorem~\ref{t:2}). As a simple corollary to this
result we obtain that the proposed transformation is \emph{non-termination preserving}
(Corollary~\ref{c:2}). In our analysis we restrict to
well-formed JBC programs that only make use of non-recursive methods.
The proposed transformation encompasses two stages. The first stage
provides a finite representation of all execution paths of $\Program$
through a graph, dubbed \emph{computation graph} (Theorem~\ref{t:1}). 
The nodes of the computation graph are abstractions of JVM states 
and the graph is formed by \emph{symbolic execution} essentially employing \emph{widening} 
akin to those used in abstract interpretations~\cite{CC:1977}. 
We develop a new graph-based representation of abstractions of JVM states
(Definition~\ref{d:state}). Furthermore we show that finiteness of the computation
graph can always be guaranteed (Lemma~\ref{l:finite}). In the second
stage, we encode the (finite) computation graph as 
\emph{constrained term rewrite system} (\emph{cTRS} for short).
CTRSs form a special type of rewrite systems that allow the formulation of conditions $C$
over a theory $\theory$, such that a rule can only be used if
the condition $C$ is satisfied in $\theory$. Constraints are used to 
express relations on program variables.

We emphasise, that the proposed transformation is not directly automatable, 
but its implementation asks for a \emph{combination} with an external shape analysis 
as presented for example in~\cite{SS05,RS06,GZ13}.
This allows the mating of the proposed term-based abstraction technique with more
standard concepts from static program analysis.
In principle, the established transformation allows for the use of
rewriting-based runtime complexity analysis for the resource analysis
of JBC programs. However, currently existing methods for complexity analysis
do not (yet) extend to cTRSs; this is subject to future work. 

\subsection{Related Work}

Our work was inspired by Panitz and Schmidt-Schauß original observation
that term-based abstraction can provide powerful termination analysis~\cite{PS97}.
Furthermore, we got inspiration from the ongoing quest to establish non-termination
preserving transformations from JBC programs to integer term rewrite
system~\cite{OBEG10,BOEG10,BMOG12b}. 
The approach has been implemented in~\aprove%
\footnote{\url{http://aprove.informatik.rwth-aachen.de/}.}
and has shown significant power in comparison to dedicated complexity 
and termination tools for JBC programs~\cite{SMP10,AAGPZ12}.
Comparing our work with earlier results 
reported for the termination graph method~\cite{OBEG10,BOEG10} we
see that a similar transformation from graphs to rewrite systems is employed.
On the other hand in Otto et al.~\cite{OBEG10} (and follow-up work)
sharing is dealt with \emph{explicitly}, while in our context sharing
is always allowed if not stated otherwise. Furthermore Otto et al.\ rely
on heuristics to obtain a finite termination graph, while we can prove
finiteness of computation graphs.

Termination behaviour and complexity of JBC programs is studied by 
Albert et al.\ in~\cite{AAGPZ12}. The approach employs program
transformations to \emph{constrained logic programs} and has been successfully
implemented in the \costa%
\footnote{\url{http://costa.ls.fi.upm.es/}.}
tool; it often allows precise bounds on the resource usage and is 
not restricted to runtime complexity.
A theoretical limitation of the work is the focus on a path-length analysis of
the heap, which does not provide the same detail as the term based abstraction
presented here.
Zuleger et al.~\cite{ZGSV:2011} employs \emph{size-change abstraction}
to analyse the runtime complexity of C programs automatically. 
In connection with pathwise analysis and \emph{contextualisation}
size-change abstraction yields a powerful analysis. The
approach has been implemented in the tool~\loopus.
Our approach extends the use of transition systems by cTRSs, which theoretically
form a strict extension. Furthermore, as our methods are rooted in rewriting we
are not limited to the powers of invariant generation tools. 
Very recently Hofmann and Rodrigues proposed in~\cite{HR13} an 
automated resource analysis based on Tarjan's amortised cost 
analysis~\cite{Tarjan:1985} for object-oriented programs.
The method is implemented in the prototype~\raja%
\footnote{\url{http://raja.tcs.ifi.lmu.de}.}.

\subsection{Structure}

This paper is structured as follows. In Sections~\ref{Preliminaries} and~\ref{Jinja}
we fix some basic notions to be used in the sequel. In particular, we 
give an overview over the Jinja programming language. Our notion of abstract states
is presented in Section~\ref{Abstraction}, while computation graphs are proposed
in Section~\ref{ComputationGraph}. Section~\ref{CTRS} introduces cTRSs and
presents the transformation from computation graphs to rewrite systems.
In Section~\ref{Implementation}
we briefly mention crucial design choices for our prototype implementation.
Finally, in Section~\ref{Conclusion} we conclude.

%% file: preliminaries.tex
\section{Preliminaries}
\label{Preliminaries}

Let $f$ be a mapping from $A$ to $B$, denoted $f:A \to B$, then 
$\dom(f) = \{ x \mid f(x) \in B \}$ and
$\range(f) = \{ f(x) \mid x \in A\}$.
Let $a \in \dom(f)$. We define:
\begin{equation*}
  f\{a \mapsto v\}(x) \defsym
  \begin{cases}
    v & \text{if $x = a$}\\
    f(x) & \text{otherwise} \tpkt
  \end{cases}
\end{equation*}

We compare partial functions with \emph{Kleene equality}:
Two partial functions $f \colon \N \to \N$ and $g \colon \N \to \N$ are equal, denoted $f \eqk g$, if for all $n \in \N$ either $f(n)$ and $g(n)$ are defined and $f(n) = g(n)$ or $f(n)$ and $g(n)$ are not defined.

We usually use square brackets to denote a list.
Further, ($\cons$) denotes the cons operator, and ($\append$) is used to denote the concatenation of two lists.

\begin{definition}
  A \emph{directed graph} $G=(\nodes{G},\suc{G},\lab{G})$ 
  over the set $\LS$ of \emph{labels} is a structure
  such that $\nodes{G}$ is a finite set, the \emph{nodes} or \emph{vertices},
  $\suc{G} \colon \nodes{G} \to \nodes{G}^{\ast}$ is a mapping that associates
  a node $u$ with an (ordered) sequence of nodes, called the \emph{successors} of $u$.
  Note that the sequence of successors of $u$ may be empty: $\suc{G}(u) = []$.
  Finally $\lab{G} \colon \nodes{G} \to \LS$ is a mapping that associates each
  node $u$ with its \emph{label} $\lab{G}(u)$.  Let $u$, $v$ be nodes
  in $G$ such that $v \in \suc{G}(u)$, then there is an \emph{edge} from
  $u$ to $v$ in $G$; the edge from $u$ to $v$ is denoted as $u \to v$.
\end{definition}
\begin{definition}
  A structure $G=(\nodes{G},\suc{G},\lab{G},\elab{G})$ is called
  \emph{directed graph with edge labels} if $(\nodes{G},\suc{G},\lab{G})$ 
   is a directed graph over the set $\LS$ and $\elab{G} \colon \nodes{G} \times \nodes{G} \to \LS$ 
  is a mapping that associates each edge $e$ with its \emph{label} $\elab{G}(e)$. 
  Edges in $G$ are denoted as $u \toss{\ell} v$, where
  $\elab{G}(u \to v) = l$ and $u,v \in \nodes{G}$. 
  We often write $u \to v$ if the label is either not important or is clear from context.
\end{definition}

If not mentioned otherwise, in the following a \emph{graph} 
is a directed graph with edge labels.
Usually nodes in a graph are denoted by $u,v, \dots$ 
possibly followed by subscripts. 
We drop the reference to the graph $G$
from $\nodes{G}$, $\suc{G}$, and $\lab{G}$, ie., 
we write $G = (\Nodes,\suc,\Lab)$ if no confusion can arise from this.
Further, we also write $u \in G$ instead of $u \in \Nodes$.

Let $G=(\Nodes,\suc,\Lab)$ be a graph and let $u \in G$. Consider
$\Succ(u) = [u_1,\dots,u_{k}]$. We call $u_i$ 
($1 \leqslant i \leqslant k$) the \emph{$i$-th successor} of $u$
(denoted as $u \suci{i} u_i$). 
If $u \suci{i} v$ for some $i$, then we simply write
$u \reach v$. A node $v$ is called \emph{reachable} 
from $u$ if $u \reachtr v$, where $\reachtr$ 
denotes the reflexive and transitive closure of $\reach$.
We write $\reachtir$ for $\reach \circ \reachtr$.
A graph $G$ is \emph{acyclic} if $u \reachtir v$ implies $u \not= v$.
We write $\subgraphAt{G}{u}$ for the subgraph of $G$ reachable from $u$.

%% file: jinja.tex
\section{Jinja Bytecode}
\label{Jinja}

In this section, we give an overview over the Jinja programming
language~\cite{KN06}. In particular we inspect the internal state of the
Jinja Virtual Machine (JVM). We expect the reader to be familiar with the Java
programming language. 

\begin{definition}
\label{d:value}
A \emph{Jinja value} can be
a Boolean of type $\tybool$, 
an (unbounded) integer of type $\tyint$, 
the dummy value $\unit$ of type $\tyunit$, 
the null reference $\mynull$ of type $\tynull$, or
a reference (or address).
\end{definition}

We usually refer to (non-null) references as addresses.
The dummy value $\unit$ is used for the evaluation of
assignments (see~\cite{KN06}) and also used in the JVM
to allocate uninitialised local variables. 
The actual type of addresses is not important and we usually identify the type
of an address with the type of the object bounded to the address.

\begin{example}
\label{ex:append}
Figure~\ref{fig:append} depicts a program defining a \ttt{List} class with the \ttt{append} method.
Deviating from the notation employed by Klein and Nipkow in~\cite{KN06}, we
present Jinja code in a Java-like syntax.
\end{example}

\begin{figure}[ht]
\center{
\begin{minipage}[t]{0.5\textwidth}
\begin{lstlisting}[firstnumber=1]       
  class List{
    List next;
    int val;

    void append(List ys){
      List cur = this;
      while(cur.next != null){
        cur = cur.next
      }
      cur.next = ys;
  }
\end{lstlisting}
\end{minipage}
}
\caption{The \ttt{append} program.}
\label{fig:append}
\end{figure}

In preparation for the sequent sections, we reflect the structure and
properties of JBC programs and the JVM.
\begin{definition}
\label{d:program}
A \emph{JBC program} $\PP$ consists of a set of \emph{class declarations}. Each class
is identified by a \emph{class name} and further consists of the name of its
direct \emph{superclass}, \emph{field declarations} and \emph{method
declarations}.
The superclass declaration is non-empty, except for a dedicated class termed \emph{Object}.
Moreover, the subclass hierarchy of $\PP$ is tree-shaped.
A field declaration is a pair of \emph{field name} and
\emph{field type}. A method declaration consists of the \emph{method name}, a
list of \emph{parameter types}, the \emph{result type} and the \emph{method
body}. 
A method body is a triple of $(mxs \times mxl \times instructionlist)$, 
where $mxs$ and $mxl$ are natural numbers denoting the maximum size of the operand 
stack and the number of local variables, not including the $this$
reference and the parameters of the method, while $instructionlist$ gives a
sequence of bytecode instructions. 
The $this$ reference can be conceived as a hidden parameter and references the object that invokes the method.
 
The set of Jinja bytecode instructions is adapted for our needs and listed in
Figure~\ref{fig:bcs}. We employ following conventions: Let $n$
denote a natural number, $i$ an integer, $v$ a Jinja value, $cn$ a class name,
and $mn$ a method name.
\begin{figure}[ht]
\begin{align*}
  \ttt{Ins} \mathrel{:=~}
  &
\jload~n\mid 
\jstore~n\mid
\jpush~v\mid
\jpop\\
  &
\mid
\jiadd\mid
\jisub\mid
\jcmpgeq\mid
\jcmpeq\mid
\jcmpneq\mid
\jand\mid
\jor\mid
\jnot \\
  &
\mid \jgoto~i\mid
\jiffalse~n\mid \\
&
\mid\jnew~cn\mid
\jgetfield~fn~cn\mid
\jputfield~fn~cn\mid
\jcheckcast~cn\mid \\
&
\mid \jinvoke~mn~n\mid
\jreturn
\end{align*}
  \caption{The Jinja bytecode instruction set.}
  \label{fig:bcs}
\end{figure}
\end{definition}

\begin{definition}
\label{d:jvmstate}
A \emph{(JVM) state} is a pair consisting of the \emph{heap} and a list of \emph{frames}. 
Let $\subclassp$ denote the strict subclass relation and $\subclass$ its reflexive closure.
%
%
A \emph{heap} is a mapping from \emph{addresses} to \emph{objects}, 
where an object is a pair $(\cname,\ftable)$ such that:
\begin{itemize}
\item $\cname$ denotes the \emph{class name}, and
\item $\ftable$ denotes the fieldtable, ie., a mapping from $(\cname',
  \fname)$ to values, where $\fname$ is a \emph{field name} and $\cname'$
  is a (not necessarily proper) superclass of $\cname$, ie., $\cname
  \subclass \cname'$.
\end{itemize}
A \emph{frame} represents the environment of a method and is a
quintuple $(\opstk,\loc,\cname,\mname,\PC)$, such that:
\begin{itemize}
\item $\opstk$ denotes the \emph{operation stack}, ie., an
array of values,
\item $\loc$ denotes the \emph{registers}, ie., 
an array of values,
\item $\cname$ denotes the \emph{class name},
\item $\mname$ denotes the \emph{method name}, and
\item $\PC$ is the \emph{program counter}.
\end{itemize}
\end{definition}

Let $\opstk$ ($\loc$) denote the operation stack (registers) of a given frame.
Typically the structure of $\loc$ is as follows: the $0^{th}$ register holds
the \emph{this}-pointer, followed by the parameters and the local variables of
the method. Uninitialised registers are preallocated with the dummy value
$\unit$. 
We denote the entries of $\opstk$ ($\loc$), by $\opstk(i)$ ($\loc(i)$) for $i
\in \N$ and write $\dom(\opstk)$ ($\dom(\loc)$) for the set of indices of the
array $\opstk$ ($\loc$). 
The collection of all stack (register) indices of a state is denoted $\stkfamily$ ($\locfamily$).
Often there is no need to separate between the local variables
of a Jinja program and the registers in a JBC program. Hence
we use registers and local variables interchangeably.
%
Observe that the domain of the fieldtable for a given object of class $\cname$ contains all
fields declared for $\cname$ together with all fields declared for superclasses of
$\cname$. Clearly the domain of the fieldtable is equal for any instance of class $\cname$.

Figure~\ref{fig:jbc} illustrates the one-step execution of the \ttt{IAdd} bytecode
instruction. We have extended the original set of instructions by some
standard operations on values, taking ideas from Jinja with Threads 
into account~\cite{Lochbihler07,Lochbihler10}.
The semantics of all employed JBC instructions can be found in the Appendix.
\begin{figure}[ht]
\footnotesize
  \begin{equation*}
  \begin{array}{lll}
   \mraisebox{\jiadd} & \infer{\jvmstate{}{\heap}{\jvmframe{(i_2 + i_1) \cons \opstk}{\loc}{\cname}{\mname}{\PC+1}\cons \myframes}}{\jvmstate{}{\heap}{\jvmframe{i_2 \cons i_1 \cons \opstk}{\loc}{\cname}{\mname}{\PC}\cons \myframes}}
   \end{array}
  \end{equation*}
  \caption{The \jiadd~ bytecode instruction.}
  \label{fig:jbc}
\end{figure}

\begin{example}
\label{ex:appendbc}  
Consider the \texttt{append} program from Example~\ref{ex:append}.
Figure~\ref{fig:listbc} depicts the corresponding bytecode program, resulting
from the compilation rules in~\cite{KN06}. In the following we name the
registers $0$,$1$, and $2$ as \emph{this}, \emph{ys}, and \emph{cur}, 
respectively.
\end{example}

\begin{figure}[ht]
\center{
\begin{minipage}[t]{.8\textwidth}
\begin{lstlisting}
Class:
 Name: List              Bytecode: 
 Classbody:               00: Load 0
  Superclass: Object      01: Store 2
  Fields:                 02: Push unit
   List next              03: Pop
   int val                04: Load 2
  Methods:                05: Getfield next List
   Method: unit append    06: Push null
    Parameters:           07: CmpNeq
     List ys              08: IfFalse 7
    Methodbody:           09: Load 2
     MaxStack:            10: Getfield next List
      2                   11: Store 2
     MaxVars:             12: Push unit
      1                   13: Pop
                          14: Goto -10
                          15: Push unit
                          16: Pop
                          17: Load 2
                          18: Load 1
                          19: PutField next List
                          20: Push unit
                          21: Return
\end{lstlisting}
\end{minipage}
}
\caption{The bytecode for the \texttt{List} program.}
\label{fig:listbc}
\end{figure}

\begin{definition}
\label{d:insType}
We extend the subclass relation to a partial order on types, denoted $\insType$.
The types of $\PP$ consists of $\{ \tybool, \tyint, \tyunit, \tynull\}$ together with all classes $cn$ defined in $\PP$.
We use $\type(v)$ to denote the type of value $v$ and $\types(\PP)$ to denote the collection of types in $\PP$.
Recall that we usually identify the type of an address with the type of the object bound to the address.
Let $t, t', cn, cn'$ be types in $\PP$.
Then $t \insType t'$ holds if $t = t'$ or
\begin{itemize}
\item $t = \tyunit$,
\item $t = \tynull$ and $t' = cn$,
\item $t = cn$, $t' = cn'$ and $cn \subclass cn'$.
\end{itemize}
The \emph{least common superclass} is the least upper bound for a set of classes $CN \subseteq \types(\PP)$ and is always defined.
\end{definition}

The bytecode verifier established in~\cite{KN06} ensures following properties:
All bytecode instructions are provided with arguments of the expected type. No
instruction tries to get a value from the empty stack, nor puts more elements
on the stack or access more registers than specified in the method. The program
counter is always within the code array of the method. All registers except
from the register storing $this$ must be first written to before
accessed.  
Furthermore the verifier ensures that for states with equal program counter the
size of the stack is of equal length. Moreover, the list of registers
is of fixed length. The compiler presented in~\cite{KN06} transforms a
well-formed Jinja program into a well-formed JBC program. A JBC program that
passes the bytecode verification is again called \emph{well-formed}.

While the set of instruction used here are a (slight) extension of
the minimalistic set considered in~\cite{KN06}, this notion
of well-formedness is still applicable, as all considered extensions
are present in Jinja with Threads~\cite{Lochbihler07,Lochbihler10}.
In the following we consider Jinja programs and JBC programs to be
well-formed. To ease readability we do not consider exception handling, that
is, an exception yields immediate termination of the program.  This is not a
restriction of our analysis, as it could be easily integrated, but
complicates matters without gaining additional insight.%

While Definition~\ref{d:jvmstate} provides a succinct presentation of the state, it is more natural to conceive the heap (and conclusively a state) as a graph.
We omit the technical definition here but provide the general idea:
Let $s = (\heap,\myframes)$ be a state.
We define the \emph{state graph} of $s$ as $\State = (\nodes{\State},\suc{\State},\lab{\State},\elab{\State})$.
For all non-address values of $s$ we define an unique \emph{implicit reference}.
The idea is that sharing is only induced via references but not implicit references.
The nodes of $S$ consists of all stack (register) indices, the references in $\heap$ and the implicit references of $s$.
The successors of a node indicate the values bound to stack (register) indices and the fields of instances in $\heap$, and is an implicit reference if a non-address value is bound and a reference otherwise.
The label of a node is either a stack (register) index, the type of an instance $\heap(u)$ or a non-address value.
The label of an edge indicates the fields $(cn,id)$ for instances $\heap(u)$, and is empty otherwise.

In presenting state graphs, we indicate references, but do not depict implicit references.
Furthermore, we use representative names for stack (register) indices.

\begin{example}
  Recall the $\ttt{append}$ program of example~\ref{ex:append}.
  Suppose $this$ is initially a list of length one, and $ys$ is $\mynull$.
  Figure~\ref{fig:Gr} depicts the state graph after the assignment $cur = this$.
\begin{figure}[ht]
      \centering
      \footnotesize
      \begin{tikzpicture}[node distance=1.5cm, descr/.style={fill=white}]
        \def\mshift{1.5cm}
        \def\mshiftx{1.15cm}
        \node (l)                            { this };
        \node (c)   at (l)  [xshift=\mshift] { cur };
        \node (y)   at (c)  [xshift=\mshift] { ys };

        \node (o1)  [below of=l]  { $o_1 \colon \m{List}$ };
        \node (o2)  [below of=o1] { $\mynull$ };
        \node (o1i) [left of=o2] { $0$ };
        \node (oy)  [below of=y] { $\mynull$ };

        \draw (l) -- (o1);
        \draw (c) -- (o1);
        \draw (y) -- (oy);
        \draw (o1) edge [->] node [descr] {\footnotesize next} (o2);
        \draw (o1) edge [->] node [descr] {\footnotesize val} (o1i);
      \end{tikzpicture}  
      \caption{State graph.}
      \label{fig:Gr}
\end{figure}
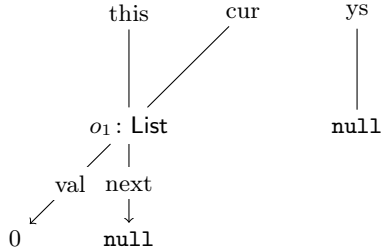
\end{example}

Let $P$ be a program and let $s$ and $t$ be states. Then we denote by $\JVMstep{s}{t}$
the one-step transition relation of the JVM. If there exists a (normal) evaluation
of $s$ to $t$, we write $\JVMexec{s}{t}$. Let $\States$ denote the set
of states. The complete lattice 
$\Pow(\States) \defsym (\Pow(\States),\subseteq,\cup,\cap,\varnothing,\States)$
denotes the \emph{concrete computation domain}.

The \emph{size} of a state is defined on a \emph{per-reference} basis, which unravels sharing.
We explicitly add $1$ to the overall construction.
This does not affect the results but allows a more convenient relation to the size of its term representation we present later.
\begin{definition}
\label{d:size} 
Let $\state$ be a state and let $\State$ be its state graph. 
Let $u,v$ be nodes in $\State$ and $u \reachtr[S] v$ denote a simple path $P$ in $\State$ from 
$u$ to $v$. Note that $P$ does not contain cycles.
Then the size of a stack or register index $u$, denoted
as $\statesize{u}$, is defined as follows:
\begin{equation*}
  \statesize{u} \defsym \sum_{\raisebox{-2mm}{$u \reachtir[S] v$}} \statesize{\lab{\State}(v)} \tkom
\end{equation*}
where $\statesize{l}$ is $\abs(l)$ if $l \in \Z$, 
otherwise $1$, for $l \in \lab{\State}$. Here, $\abs(z)$ denotes the absolute value of the integer $z$.
Then the \emph{size} of $\state$ is the sum of 
all sizes of stack or register indices in $\State$ plus $1$.
In the following we use $\statesize{s}$ to denote the size of a state $s$.
\end{definition}

We define the \emph{runtime} of a JVM for a given normal evaluation $\JVMexec{s}{t}$
as the number of single-step executions in the course
of the evaluation from $s$ to $t$. 
\begin{definition}
  Let $\States$ denote the set of JVM states of $\PP$, and $\mathcal{S} \subseteq \States$.
We define the \emph{runtime complexity} with respect to $\Program$ as follows:
\[  
\rcjvm(n) \eqk
\max \{ m | 
\text{$\JVMexec{\start}{t}$ holds such that the runtime is $m$,
$\start \in \mathcal{S}$  and $\statesize{\start} \leqslant n$
}\} \tpkt
\]
Note that we adopt a (standard) unit cost model for system calls.
\end{definition}

%% file: abstraction.tex
\section{Abstract States}
\label{Abstraction}

In this section, we introduce \emph{abstract states} as generalisations
of JVM states. The intuition being
that abstract states represent sets of states in the JVM. The idea of
abstracting JVM states in this way is due to Otto et al.~\cite{OBEG10}. However,
our presentation crucially differs from~\cite{OBEG10} (and also from follow-up work in the
literature) as we employ an implicit representation of sharing that makes use of
graph morphisms, rather than the explicit sharing information proposed 
in~\cite{OBEG10,BOEG10,BOG11,BMOG12b}. Furthermore, abstract states as defined below are
a straightforward generalisation of JVM states as defined in~\cite{KN06}. 
This circumvents an additional transformation step as presented  in~\cite{BOEG10}.

\begin{definition}
We extend Jinja expressions by countable many 
abstract variables $X_1,X_2,X_3,\dots$, denoted by $x$, $y$, $z$, \dots 
An \emph{abstract variable} may either abstract an object,
an integer or a Boolean value. 
\end{definition}

In denoting abstract variables typically 
the name is of less importance than the type, that
is we denote an abstract variable for an object of class $\cname$, simply as
$\AbsClass$, while abstract integer or Boolean variables are denoted as $\AbsInt$, 
and $\AbsBool$, respectively. The (strict) subclass relation ($\subclassp$) $\subclass$ 
is extended in the natural way to abstract variables for classes.
For brevity we sometimes refer to an abstract variable of integer or Boolean type, 
as \emph{abstract integer} or \emph{abstract Boolean}, respectively.

\begin{definition}
An \emph{abstract value} is either a Jinja value (cf.~Definition~\ref{d:value}),
or an abstract Boolean or integer. In turn a Jinja value is also called a
\emph{concrete value}.
\end{definition}

Note that, as in the JVM, only (abstract) objects can be shared. 
In particular abstract variables for objects are only 
referenced via the heap. The next definition abstracts the heap of a JVM through the use
of abstract variables and values.
\begin{definition}
  \label{d:aheap}
  An \emph{abstract heap} is a mapping from \emph{addresses} to 
  \emph{abstract objects}, where an abstract object is either a pair
  $(\cname,\ftable)$ or an abstract variable.
  \emph{Abstract frames} are defined like frames of the JVM, but 
registers and operand stack of an abstract frame store abstract values.
\end{definition}
We define (partial) projection functions $\classof$ and $\ftof$ as follows:
\begin{align*}
  \classof(obj) & \defsym
  \begin{cases}
    \cname & \text{if $obj$ is an object and $obj = (\cname,\ftable)$} \\
    \AbsClass & \text{if $obj$ is an abstract variable of type $\cname$}
  \end{cases}
\\
  \ftof(obj) & \defsym 
  \begin{cases}
    \ftable & \text{if $obj$ is an object and $obj = (\cname,\ftable)$} \\
    \text{undefined} & \text{otherwise}
  \end{cases}
\tpkt
\end{align*}
%
%
Furthermore, we define \emph{annotations} of addresses
in an abstract state $s$, denoted as $\isunshared$. Formally,
annotations are pairs $\unshare{p}{q}$ of addresses, 
where $p,q \in \heap$ and $p$ is not $q$.

\begin{definition}
\label{d:state}
An \emph{abstract state} $s=(\heap,\myframes,\isunshared)$ 
is either a triple consisting of an abstract heap $\heap$, 
a list of abstract frames $\myframes$, and a set of annotations
$\isunshared$, the \emph{maximal abstract state}, denoted as $\top$, 
or the \emph{minimal abstract state}, denoted
as $\bottom$.
If $s=(\heap,\myframes,\isunshared)$, we demand that all addresses in $\heap$
are reachable from local variables or stack entries in the list
of frames $\myframes$. 
The set of abstract states is collected in the set $\AStates$.
\end{definition}

When depicting (abstract) states, we replace stack and register indices by
intuitive names, denoted in roman font. Furthermore, we make use of the following
conventions: we use an italic font (and lower-case) to describe abstract variables and
a sans serif (and upper-case) to depict class names. 
\begin{example}
\label{ex:A}  
Consider the \ts{List} program from Example~\ref{ex:append} together
with the well-formed JBC program depicted in Figure~\ref{fig:listbc}. 
Consider the state $A$ depicted below:
\begin{center}
\begin{tikzpicture}
\footnotesize
\node[state](A){
      \parbox{.45\textwidth}{
        $
        \begin{array}[ht]{l|l}
          \m{04} & \epsilon \mid \text{this} = o_1, \text{ys} = o_2, \text{cur} = o_1 \\
          & o_1 = \m{List} (\text{List.val} = int, \text{List.next} = o_3 ) \\
          A & o_2 = list, o_3 = list 
        \end{array}
        $
      }
    };  
\end{tikzpicture}
\end{center}
The operation stack in $A$ is empty. The registers \emph{this} and \emph{cur}
contain the same address $o_1$ and \emph{ys} is mapped to $o_2$. 
In the heap $o_1$ is mapped to an object of type $\m{List}$ whose
value is abstracted to $int$ and whose 
next element is referenced by $o_3$.
It is not difficult to see that $A$ forms an abstraction of any
JVM state obtained at instruction $\m{04}$ in the \ts{List} program
(if \emph{this} initially references a non-empty list)
before any iteration of the \ttt{while}-loop.
Furthermore, consider the following state $B$:
\begin{center}
\begin{tikzpicture}
\footnotesize
\node[state](B){
      \parbox{.45\textwidth}{
        $
        \begin{array}[ht]{l|l}
          \m{04} & \epsilon \mid \text{this} = o_1, \text{ys} = o_2, \text{cur} = o_3  \\
          & o_1 = \m{List}(\text{List.val} = int, \text{List.next} = o_3) \\
          & o_2 = list, o_4 = list  \\
          B & o_3 = \m{List}(\text{List.val} = int, \text{List.next} = o_4 )
        \end{array}
        $
      }
    };    
\end{tikzpicture}
\end{center}
Again it is not difficult to see that $B$ abstracts any JVM state obtained
if exactly one iteration of the loop has been performed.
\end{example}

Due to the presence of abstract variables, abstract states can represent sets
of states as the variables can be suitably instantiated. The annotation
$\unshare{p}{q} \in \isunshared$ will be used to disallow aliasing of addresses
in JVM states represented by the abstract state. Different
JVM states can be abstracted to a single abstract state. To make this
precise, we will augment $\AStates$ with a partial order $\instance$, the
\emph{instance relation} (see Definition~\ref{d:instance2}). We will extend
the partial order $(\AStates,\instance)$ to a complete lattice 
$\AStates \defsym (\AStates,\instance,\join,\meet,\bottom,\top)$ and 
show a Galois insertion between $\Pow(\States)$ and $\AStates$.
%
%

\begin{definition}
\label{d:instance3}
We define a preorder on abstract values, which are not references, 
and abstract objects.
We extend $\type(v)$ (cf. Definition~\ref{d:insType}) to abstract values the intended way, ie., $\type(int) = int, \type(bool) = bool$ and $\type(cn) = cn$ for an integer variable $int$, a Boolean variable $bool$, and class variable $cn$.
Then the preorder $\insBasic$ is defined as follows:
We have $v \insBasic w$, if either 
\begin{enumerate}
\item $v = w$, or
\item $\type(v) \insType \type(w)$ and $w$ is an abstract variable.
\end{enumerate}
We write $w \absBasic v$, if $v \insBasic w$.
\end{definition}





Let $\card{\opstk}$, $\card{\loc}$ denote the maximum size of the operand stack
and the number of variables respectively. We make use of
the following abbreviation: $w \absAux{m} v$ if either
$w \absBasic v$ or $v,w$ are references and we have
$v = m(w)$, where $m$ denotes a mapping on references.

\begin{definition}
\label{d:instance2}
Let $s = (\heap,\myframes,\isunshared)$ be
a state in $\AStates \setminus \{\top,\bottom\}$ with 
$\myframes = [\myframe_1,\dots,\myframe_k]$
and $\myframe_i = (\opstk_i,\loc_i,\cname_i,\mname_i,\PC_i)$, and
let $t=(\heap',\myframes',\isunshared')$ be a state with 
$\myframes' = [\myframe'_1,\dots,\myframe'_k]$
and $\myframe'_i = (\opstk'_i,\loc'_i,\cname'_i,\mname'_i,\PC'_i)$.
Then $s$ is an \emph{abstraction} of $t$ (denoted as $s \abstraction t$) if
the following conditions hold:
\begin{enumerate}
\item for all $1 \leqslant i \leqslant k$: $\PC_i = \PC'_i$, 
$\cname_i = \cname'_i$, and $\mname_i = \mname'_i$,
\item for all $1 \leqslant i \leqslant k$: $\dom(\opstk_i) = \dom(\opstk'_i)$
and $\dom(\loc_i) = \dom(\loc'_i)$, and
\item there exists a mapping $m \colon \dom(\heap) \to \dom(\heap')$ such that
  \begin{itemize}
  \item for all $1 \leqslant i \leqslant k$, $1 \leqslant j \leqslant \card{\opstk_i}$: 
    ${\opstk_i(j) \absAux{m} \opstk'_i(j)}$,
  \item for all $1 \leqslant i \leqslant k$, $1 \leqslant j \leqslant \card{\loc_i}$:
    $\loc_i(j) \absAux{m} \loc'_i(j)$,
  \item for all $a \in \dom(\heap)$: 
    $\heap(a) \absBasic heap'(m(a))$,
  \item for all $a \in \dom(\heap)$, such that $\ftof(\heap(a))$ is defined and
    for all $1 \leqslant i \leqslant \ell$: 
    $f(\cname_i,id_i) \absAux{m} f'(\cname'_i,id_i)$,where\\
    $f \defsym \ftof(\heap(a))$ with
    $\dom(f) = \{(\cname_1,id_1),\dots,(\cname_\ell,id_\ell)\}$, and\\
    $f' \defsym \ftof(\heap'(m(a)))$ with
    $\dom(f') = \{(\cname_1,id_1),\dots,(\cname_\ell,id_\ell)\}$.
  \end{itemize}
\item finally, we have $\isunshared' \supseteq m^\ast(\isunshared)$.
\end{enumerate}
Here, $m^\ast$ denotes the lifting of the mapping $m$ to sets:
$m(\{isunshared_1,\dots,\isunshared_k\}) = \{m(\isunshared_1),\dots,m(\isunshared_k)\}$.
Furthermore for all $s \in \AStates$: $s \instance \top$ and $\bottom \instance s$.
\end{definition}

\begin{example}
\label{ex:S}
Consider the states $A$ and $B$ described in Example~\ref{ex:A}.
For the state $S$ depicted below we obtain that $A \instance S$ 
and $B \instance S$, ie., $S$ forms an abstraction of both states.
\begin{center}
\begin{tikzpicture}
\footnotesize
    \node[state](S){
      \parbox{.45\textwidth}{
        $
        \begin{array}[ht]{l|l}
          \m{04} & \epsilon \mid \text{this} = o_1, \text{ys} = o_2, \text{cur} = o_4 \\
          & o_1 = \m{List}(\text{List.val} = int, \text{List.next} = o_3 ) \\
          & o_2 = list, o_3 = list, o_5 = list\\
          S & o_4 = \m{List}(\text{List.val} = int, \text{List.next} = o_5)
        \end{array}
        $
      }
    };    
\end{tikzpicture}
\end{center}
\end{example}

The definition of state graphs naturally extends to abstract states, when incorporating $\isunshared_{\state}$ and considering abstract values.
Furthermore, we use $\top$ to denote the state graph of $\top \in \AStates$ and the empty graph to denote $\bot \in \AStates$.

%
%
\begin{example}
\label{ex:AG}
Consider the states $A$, $B$, and $S$ presented in Examples~\ref{ex:A} and~\ref{ex:S}.
The state graph of $A$ and $B$ are given in Figure~\ref{fig:A} and
Figure~\ref{fig:B}, respectively. The state graph of the abstraction
$S$ is depicted in Figure~\ref{fig:S}.
\end{example}

\begin{figure}[ht]
    \begin{tabular}{@{}l@{\hspace{5ex}}r}
    \begin{minipage}[t]{0.40\textwidth}
      \centering
      \footnotesize
      \begin{tikzpicture}[node distance=1.5cm, descr/.style={fill=white}]
        \def\mshift{1.5cm}
        \def\mshiftx{1.15cm}
        \node (l)                            { this };
        \node (c)   at (l)  [xshift=\mshift] { cur };
        \node (y)   at (c)  [xshift=\mshift] { ys };

        \node (o1)  [below of=l]  { $o_1 \colon \m{List}$ };
        \node (o2)  [below of=o1] { $o_2 \colon list$ };
        \node (o1i) [left of=o2] { $int$ };
        \node (oy)  [below of=y] { $o_3 \colon list$ };

        \draw (l) -- (o1);
        \draw (c) -- (o1);
        \draw (y) -- (oy);
        \draw (o1) edge [->] node [descr] {\footnotesize next} (o2);
        \draw (o1) edge [->] node [descr] {\footnotesize val} (o1i);
      \end{tikzpicture}  
      \caption{Abstract State $A$}
      \label{fig:A}
    \end{minipage}
    &
    \begin{minipage}[t]{0.40\textwidth}
      \centering
      \footnotesize
      \begin{tikzpicture}[node distance=1.5cm, descr/.style={fill=white}]
        \def\mshift{1.5cm}
        \def\mshiftx{1.15cm}
        \node (l)                            { this };
        \node (c)   at (l)  [xshift=\mshift] { cur };
        \node (y)   at (c)  [xshift=\mshift] { ys };

        \node (o1)  [below of=l]  { $o_1 \colon \m{List}$ };
        \node (o2)  [below of=o1] { $o_2 \colon \m{List}$ };
        \node (o1i) [left of=o2] { $int$ };
        \node (o3)  [below of=o2] { $o_4 \colon list$ };
        \node (o2i) [left of=o3] { $int$ };
        \node (oy)  [below of=y] { $o_3 \colon list$ };

        \draw (l) -- (o1);
        \draw (c) edge [-,bend left=15] (o2);
        \draw (y) -- (oy);
        \draw (o1) edge [->] node [descr] {\footnotesize next} (o2);
        \draw (o1) edge [->] node [descr] {\footnotesize val} (o1i);
        \draw (o2) edge [->] node [descr] {\footnotesize next} (o3);
        \draw (o2) edge [->] node [descr] {\footnotesize val} (o2i);
      \end{tikzpicture}  
      \caption{Abstract State $B$}
      \label{fig:B}
    \end{minipage}
    \end{tabular}
\end{figure}

We introduce \emph{state homomorphisms} that allow an alternative, but
equivalent definition of the instance relation $\instance$.
\begin{definition} 
\label{d:morphism}
Let $S$ and $T$ be state graphs of states $s$ and $t$, respectively
such that $S, T \not= \varnothing$. A \emph{state homomorphism} from $S$ to $T$ 
(denoted $m \colon S \to T$) is a function 
$m \colon \nodes{S} \to \nodes{T}$ such that 
\begin{enumerate}
  \item for all $u \in S$ and $u \in \stkfamily \cup \locfamily$, 
    $\lab{S}(u) = \lab{T}(m(u))$,
  \item for all $u \in S \setminus (\stkfamily \cup \locfamily)$, 
    $\lab{S}(u) \absBasic \lab{T}(m(u))$,
 \item for all $u \in S$: if $u \suci[S]{i} v$, then $m(u) \suci[T]{i} m(v)$ and
  \item for all ${u \toss{\ell} v} \in S$ and ${m(u) \toss{\ell'} m(v)} \in T$,
$\ell = \ell'$.
\end{enumerate}
\end{definition}


If no confusion can arise we refer to a state homomorphism simply as
\emph{morphism}. It is easy to see that the composition $m_1 \circ m_2$ of
two morphisms $m_1$, $m_2$ is again a morphism. 
We say that two states $s, t \in \AStates$ are \emph{isomorphic} if there exists
a morphism from $s$ to $t$ and vice versa. Suppose the abstract states 
$s$ and $t$ are isomorphic. Then they differ only in their abstract variables and
can be transformed into each other through a renaming of variables. 
Thus the set of JVM states represented by $s$ and $t$ is equal;
we call $s$ and $t$ \emph{equivalent} (denoted $s \equivalent t$).

Let $\state, t \in \AStates$ and let $\State$ and $T$ denote their state graphs.
Then $s \abstractionGraph t$ if one of the following alternatives holds:
(i) $S = \top$, (ii) $T$ is empty, or 
(iii) $S,T \not= \varnothing$ and there exists a state morphism $m$ from $S$ to $T$; 
$s=(\heap,\myframes,\isunshared)$, $t=(\heap',\myframes',\isunshared')$ 
and the program counters, the class and method names of all frames in $s$ and $t$ coincide;
$\isunshared' \supseteq m^\ast(\isunshared)$. 

\begin{lemma}
\label{l:instance}
Let $\state, t \in \AStates$. Then $s \instance t$ iff $s \instanceGraph t$.
\end{lemma}
\begin{proof}
Straightforward.  
\end{proof}

Due to Lemma~\ref{l:instance} and the composability of 
morphism it follows that the instance relation $\instance$ is transitive. 
Hence the relation $\instance$ is a preorder.
Furthermore $\instance$ can be lifted to a partial order, if we consider the
factorisation of the set of abstract states with respect to the equivalence
relation $\equivalent$. In order to express this fact notationally, we identify 
isomorphic states and replace $\equivalent$ by $=$. 
Conclusively $(\AStates,\instance)$ is a partial order.
We are left to provide a least upper bound definition of the \emph{join} of 
abstract states. 
\begin{definition}
\label{d:abstraction}
Let $s$ and $s'$ be states such that there exists an abstraction $t$ of
$s$ and $s'$. We call $t$ the \emph{join} of $s$ and $s'$, denoted
as $s \join s'$, if $t$ is a least upper bound of $\{s,s'\}$ with
respect to the preorder $\instance$.
\end{definition}

The limit cases are handled as usual.
If the program locations of $s$ and $s'$ differ, then  $s \join s' = \top$.
Otherwise, we can identify invariants to construct an upper bound $t \neq \top$ and prove well-definedness of $s \join s'$.
Let $\State=(\nodes{\State},\suc{\State},\lab{\State},\elab{\State},\isunshared_{\State})$ 
and $\State'=(\nodes{\State'},\suc{\State'},\lab{\State'},\elab{\State'},\isunshared_{\State'})$ 
be the two state graphs of state $s$ and $s'$, respectively.
Furthermore, let $t$ be an abstraction of $s$ and $s'$, and let
$T=(\nodes{T},\suc{T},\lab{T},\elab{T},\isunshared_{T})$ be its state graph. 
By definition we have the following properties:
\begin{enumerate}
\item \label{en:abs:i}
Let $\stkfamily$ ($\locfamily$) collect the stack (register) indices
  of state $s$. As $s \instance t$, $\stkfamily$ ($\locfamily$) coincides
  with the set of stack (register) indices of $t$. Similarly for $s'$ and
  thus $\nodes{T} \supseteq {\stkfamily \cup \locfamily}$.
\item \label{en:abs:ii}
  For any node $u \in T$ there exist uniquely defined nodes $v \in \nodes{\State}$,
  $w \in \nodes{\State'}$ such that $\lab{\State}(v) \insBasic \lab{T}(u)$,
  $\lab{\State'}(w) \insBasic \lab{T}(u)$. We say the nodes $v$ and $w$
  \emph{correspond} to $u$.
\item \label{en:abs:iii}
  For any node $u \in T$ and any successor $u'$ of $u$ in $T$
  there exists a successor $v'$ ($w'$) in $\State$ ($\State'$) of
  the corresponding node $v$ ($w$) in $\State$ ($\State'$). Furthermore
  $v'$ and $w'$ correspond to $u'$.
\item \label{en:abs:iv}
  For any edge ${u \toss{\ell} u'} \in T$ such that $v$ ($w$) 
  corresponds to $u$ in $\State$ ($\State'$) there is an
  edge ${v \toss{k} v'} \in \State$ and an 
  edge ${w \toss{k'} w'} \in \State'$ such that $\ell = k = k'$.
\item \label{en:abs:v}
  For any annotation ${u \not = u'} \in \isunshared_{T}$ there
  exists ${v \not = v'}$ in $\isunshared_{\State}$ and 
   ${w \not = w'}$ in $\isunshared_{\State'}$, where $v$ ($v'$) and $w$ ($w'$)
   correspond to $u$ ($u'$).
\end{enumerate}

In order to construct an abstraction $t$ of $s$ and $s'$
we use the above properties as invariants and define its state graph $T$ by iterated
extension. 
We define $T^0$ by setting $\nodes{T^0} \defsym \stkfamily \cup \locfamily$. 
Due to Property~\ref{en:abs:i} these nodes exist in $\State$ and $\State'$ as well. The
labels of stack or register indices trivially coincide in $\State$ and $\State'$,
cf.~Definition~\ref{d:morphism}. Thus we set $\lab{T^0}$ accordingly. Furthermore
we set $\suc{T^0} = \elab{T^0} = \isunshared_{T^0} \defsym \varnothing$. Then
$T^0$ satisfies Properties~\ref{en:abs:i}--\ref{en:abs:v}. 

Suppose state graph $T^n$ has already been defined such that 
the \mbox{Properties~\ref{en:abs:i}--\ref{en:abs:v}} are fulfilled. 
In order to update $T^n$, let $u \in \nodes{T^n}$ such that
$v$ and $w$ correspond to $u$. Suppose ${v \toss{k} v'} \in \State$ 
and ${w \toss{k} w'} \in \State'$ such that there is no node $u'$ in $T^n$
where $v'$ and $w'$ correspond to $u'$. 
Let $u'$ denote a node fresh to $T^n$. 
We define $\nodes{T^{n+1}} \defsym \nodes{T^{n}} \cup \{u'\}$ and
establish Property~\ref{en:abs:ii} by setting $\lab{T^{n+1}}(u')$
such that $\lab{\State}(v') \insBasic \lab{T^{n+1}}(u')$
and $\lab{\State'}{w'} \insBasic \lab{T^{n+1}}(u')$ where
$\lab{T^{n+1}}(u')$ is as concrete as possible. 
If we succeed, we fix that $v'$ and $w'$ correspond to $u'$. 
It remains to update $\isunshared_{T^{n+1}}$ suitably such that 
Property~\ref{en:abs:v} is fulfilled. If this also succeeds
\mbox{Properties~\ref{en:abs:i}--\ref{en:abs:v}} are fulfilled for $T^{n+1}$. 
On the other hand, if no further update is possible we set $T \defsym T^n$. By construction
$T$ is an abstraction of $S$ and $S'$ and indeed represents $s \join s'$.

\begin{example}
\label{ex:S2}
Consider the states $A$, $B$, and $S$ described in Example~\ref{ex:AG}.  In
Figure~\ref{fig:S} an abstraction of $A$ and $B$ is given. In particular,
abstraction $S$ results of the construction defined above, ie., $S = A \join B$.

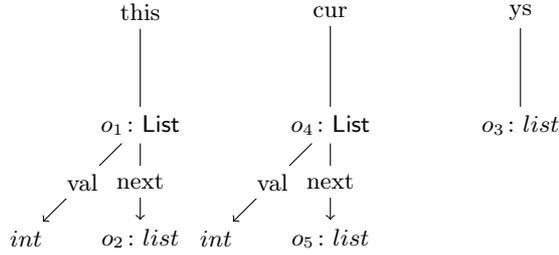
\begin{figure}[ht]
  \centering
  \begin{minipage}[b]{.55\linewidth}
  \footnotesize
  \begin{tikzpicture}[node distance=1.5cm, descr/.style={fill=white}]
    \def\mshift{4.5cm}
    \def\mshiftx{2.5cm}
    \node (l)                            { this };
    \node (c)   at (l)  [xshift=\mshiftx] { cur };
    \node (y)   at (c)  [xshift=\mshiftx] { ys };

    \node (o1)  [below of=l]  { $o_1 \colon \m{List}$ };
    \node (o2)  [below of=o1] { $o_2 \colon list$ };
    \node (o1l) [left of=o2] { $int$ };
    \node (o4)  [below of=c]  { $o_4 \colon \m{List}$ };
    \node (o5)  [below of=o4] { $o_5 \colon list$ };
    \node (o4l) [left of=o5] { $int$ };
    \node (oy)  [below of=y] { $o_3 \colon list$ };

    \draw (l) -- (o1);
    \draw (c) -- (o4);
    \draw (y) -- (oy);
    \path (o1) edge [->] node [descr] {\footnotesize next} (o2);
    \path (o1) edge [->] node [descr] {\footnotesize val} (o1l);
    \path (o4) edge [->] node [descr] {\footnotesize next} (o5);
    \path (o4) edge [->] node [descr] {\footnotesize val} (o4l);
  \end{tikzpicture}  
  \caption{Abstraction $S$}
  \label{fig:S}
  \end{minipage}%
\end{figure}
\end{example}

A sequence of states $(s_i)_{i \geqslant 0}$ forms an ascending sequence, if
$i < j$ implies $s_i \instance s_j$. An ascending sequence $(s_i)_{i \geqslant 0}$
eventually stablises, if there exits $i_0 \in \N$ such that for all $i \geqslant i_0$: 
$s_i = s_{i_0}$. The next lemma shows that any ascending sequence eventually stabilises.

\begin{lemma}
\label{l:chain}
The partial order $(\AStates,\instance)$ satisfies the ascending chain condition,
that is, any ascending chain eventually stabilises. 
\end{lemma}
\begin{proof}
In order to derive a contradiction we assume the existence of an
ascending sequence $(s_i)_{i \geqslant 0}$ that never stabilises. By definition
for all $i \geqslant 0$: $\statesize{s_i} \geqslant \statesize{s_{i+1}}$. By assumption there 
exists $i \in \N$ such that for all $j > i$: $\statesize{s_i} = \statesize{s_j}$  
and $s_{i} \properinstance s_{j}$. The only possibility for two different states
$s_{i}, s_j$ of equal size that $s_i \instance s_j$ holds, is that addresses shared
in $s_i$ become unshared in $s_j$. Clearly this is only possible for a finite amount
of cases. Contradiction.
\end{proof}

Lemma~\ref{l:chain} in conjunction with the fact that $(\AStates,\instance)$ has 
a least element $\bottom$ and binary least upper bounds implies that 
$(\AStates,\instance,\lub,\glb,\bottom,\top)$ is a complete lattice.
In particular any set of states $\mathcal{S}$ has a
least upper bound, denoted as $\lub{\mathcal{S}}$.
The meet operation $\glb{}$ can be expressed by $\lub{}$, yet in practice we do not need it.

\subsection{Correctness}

In the remainder of the paper we fix to a concrete
JBC program $\Program$. 
Above, we already restricted our attention to well-formed JBC programs $\Program$ using
the expressions and instructions defined in Section~\ref{Jinja}. For the proposed 
static analysis of these programs we additionally restrict to \emph{non-recursive} methods.
Note that the states in $\AStates$ can in principle express recursive methods, but 
for recursive methods, we cannot use the below proposed construction
to obtain \emph{finite} computation  graphs, as the graphs defined in
Definition~\ref{d:computationgraph} cannot handle unbounded list of frames.
In the following we use superscript~$\natural$, if we want to distinguish
between concrete and abstract states, or between operations on concrete and
abstract states.


%


Let $s=(\heap,\myframes) \in \States$, we define a mapping $\beta \colon \States \to \AStates$,
that injects JVM states into $\AStates$. For that let $\dom(\heap) = \{p_1,\dots,p_n\}$
and define $\isunshared$ such that all $p_i \not= p_j \in \isunshared$ for
all different $i,j$. 

\begin{definition}
We define the \emph{abstraction function} $\alpha \colon \Pow(\States) \to \AStates$ and
the \emph{concretisation function} $\gamma \colon \AStates \to \Pow(\States)$ as follows:
\begin{align*}
  \alpha(\mathcal{S}) &\defsym \lub{\{\beta(s) \mid s \in \mathcal{S}\}} \tkom \\
  \gamma(\as) &\defsym \{ s \in \States \mid \beta(s) \instance \as\} \tpkt
\end{align*}
We set $\alpha(s) \defsym \alpha(\{s\})$.
\end{definition}

It is easy to see that $\AStates$ contains redundant states:
Consider abstract states $\as,\at \in \AStates$.
Let $\as = (\heap, \myframes, \isunshared)$, $p,q \in \dom(\heap)$ and $p \neq q \in \isunshared$.
Let $\at$ be defined like $\as$ but $p \neq q \notin \isunshared$.
Now suppose that the types of $p$ and $q$ are not related with respect to the subclass order.
Then $\as \properinstance \at$ and $\gamma(\as) = \gamma(\at)$.
To form a Galois insertion between $\Pow(\States)$ and $\AStates$, we introduce a reduction operator that adds annotations for non-aliasing addresses.

\begin{definition}
\label{d:redop}
Let $\as = (\heap, \myframes, \isunshared)$ be an abstract state.
We define the \emph{reduction operator} $\varsigma \colon \AStates \to \AStates$ as follows:
\[
\varsigma(\as) \defsym (\heap, \myframes, \isunshared')\tkom
\]
where $\isunshared' \defsym \{ p \neq q \mid p, q \in \dom(heap)\} \setminus \{ p \neq q \mid s \in \gamma(\as), m: \as \to \beta(s), m(p) = m(q) \}$.
Then $\varsigma(\as) \instance \as$ and $\gamma(\varsigma(\as)) = \gamma(\as)$.
\end{definition}
In practice, we compute the reduction by a unification argument of $p$ and $q$ in $\as$:
We try to construct a new state ${\at} \instance \as$, where $r = m(p) = m(q)$.
Let $\aT$ and $\aS$ be the state graphs of $\at$ and $\as$.
Suppose $u, v, w$ represent $r, p, q$ in $\aT$ and ${\aS}$.
We can use a similar reasoning we used for the join construction, but now require $\lab{\aT}(u) \insBasic \lab{\aS}(v)$  and $\lab{\aT}(u) \insBasic \lab{\aS}(w)$ if $v$ and $w$ correspond to $u$.
If the construction succeeds, we can easily find a concrete state from ${\at}$ such that $m(p) = m(q)$.
The construction does not succeed if, for example, successors of corresponding nodes have  different concrete values; then we add $p \neq q$.

\begin{lemma}
The maps $\alpha$ and $\gamma$ define a Galois insertion between
the complete lattices $\Pow(\States)$ and
$\varsigma^\ast(\AStates)$, where $\varsigma^\ast$ denotes the set extension of $\varsigma$.
\end{lemma}
\begin{proof}
  It suffices to prove that $\gamma$ is injective, ie., for all $\as, \at \in \varsigma^\ast(\AStates)$ if $\as \neq \at$ then $\gamma(\as) \neq \gamma(\at)$.
  Suppose $\as \neq \at$ but $\gamma(\as) = \gamma(\at)$.
  It is a simple consequence of our morphism definition that $\gamma(\as) \neq \gamma(\at)$, if the state graphs of $\as$ and $\at$ differ.
  Hence, $\as$ can only be different from $\at$ if the annotations of $\as$ and $\at$ differ.
  However, by assumption they are equal.
  Contradiction.  
\end{proof}

It follows that the reduction operator defined in Definition~\ref{d:redop}, indeed returns the greatest lower bound that represents the same element in the concrete domain as required.
In the following we identify the $\varsigma^\ast(\AStates)$ with $\AStates$.

In order to prove that the abstract domain $\AStates$ correctly approximates 
the concrete domain $\Pow(\States)$ we need to define a suitable notion
of \emph{abstract computation} on abstract states. 
Recall that Figure~\ref{fig:jbc} presents the single-step execution of 
the $\jiadd$ instruction on the JVM. Based on these instructions, and
actually mimicking them quite closely, we define how abstract states are
evaluated symbolically. This is straightforward in most cases, with the
exception of $\jputfield$ and $\jcmpeq$ instructions.
With respect to the former, we suppose a preliminary analysis on different heap shape properties.
In particular our analysis requires \emph{may-share}, \emph{may-reachable}, and 
\emph{maybe-cyclic} analyses as given, see for example~\cite{SS05,RS06,GZ13}. 
%
\begin{definition}
\label{d:share}
  Let $\as$ be an abstract state and $p,q$ be addresses in the heap of $s$.
  We use $S$ to denote the state graph of $\beta(s)$ for some concrete state $s$.
  We say that:
  \begin{itemize}
    \item $p$ and $q$ \emph{may-alias},   if $m(p) = m(q)$ for some $s \in \gamma(\as)$ and morphism $m \colon \as \to \beta(s)$;
    \item $p$ \emph{may-reaches} $q$,     if $m(p) \reachtr[S] m(q)$ for some $s \in \gamma(\as)$ and morphism $m \colon \as \to \beta(s)$;
    \item $p$ is \emph{maybe-cyclic},     if $m(p) \reachtir[S] m(p)$ for some $s \in \gamma(\as)$ and morphism $m \colon \as \to \beta(s)$;
    \item $p$ is \emph{acyclic},          if $p$ is not maybe-cyclic.
  \end{itemize}
\end{definition}
Note that our representation does not provide a precise approximation of these properties, as abstract variables generally also present cyclic instances.
%

%

In Figure~\ref{fig:symeval} we have worked out the cases for the
instructions $\h\jload$, $\h\jiadd$, $\h\jcmpeq$, $\h\jiffalse$, $\h\jnew$ and $\h\jputfield$.
We follow the notation used in Figure~\ref{fig:jbc} above.
The other cases are left to the reader.
In addition to symbolic evaluations, we define refinement steps on 
abstract states $\as$ if the information given in $\as$ is not concrete enough
to execute a given instruction. It will be a consequence of our definitions that 
for any refinement $\as_i$ of $\as$, we have $\as_i \instance \as$.

In the following assume $\as = (\heap,\myframes,\isunshared)$.
Some comments: 
The symbolic instruction $\h{\jload}~n$ loads the value of the $n$th register onto the stack.
The only difference to $\jload~n$ is that the value may be an integer or Boolean variable.
For the $\h\jiadd$ instruction, we introduce a new abstract integer $i_3$
and the side-condition $i_1 + i_2 = i_3$, if either $i_1$ or $i_2$ is an integer variable.
The $\h\jcmpeq$ splits into different cases, depending on the status of the
compared values. We adapt the instruction to abstract values
as follows:
\begin{enumerate}
\item Let $val_1$ and $val_2$ be addresses. If the addresses of $val_1$ and
$val_2$ are the same then the test evaluates to \ts{true}. Otherwise, we have
to check if $val_1$ and $val_2$ may alias and perform a unsharing refinement 
(cf.~Definition~\ref{d:unsharing}) if necessary. In the latter
case the test returns \ts{false}.
\item Wlog.\@ let $val_1$ be an address and $val_2$ be $\mynull$. If
$\heap(val_1)=obj$ and $\classof(obj) = \cname$, we perform a instance refinement according to
Definition~\ref{d:classinstance}  on $val_1$ and re-consider the condition.
\item If $val_1$ and $val_2$ are concrete non-address Jinja values, then the test $(val_1 = val_2)$ can be directly executed and the symbolic execution equals the instruction on the JVM.
\item If $val_1$ and $val_2$ are abstract Boolean or integer variables, then we introduce a new Boolean variable $b_3$ and the side condition $(val_1 = val_2) \equiv b_3$.
Figure~\ref{fig:symeval} only shows the latter case.
\end{enumerate}
$\h\jnew~cn$ allocates a new instance of type $cn$ in the heap and pushes the corresponding address onto the stack.  
All fields of the fresh created instance are instantiated with the default value.
That is, $0$ for integer typed fields, $\false$ for Boolean typed fields, and $\mynull$ otherwise.
%
If the top element of the stack is a concrete value, $\h\jiffalse$ can be executed directly.
Otherwise we perform a \emph{Boolean refinement}, replacing the variable with values $\true$ and $\false$.
Recall that a class variable $cn$ represents $\mynull$ as well as instances of $cn$ and its subtypes.
Hence, $\h{\jputfield}~\fname~\cname'$ may require an instance refinement (cf.~Definition~\ref{d:classinstance}).
Let $v$ be a value and $p$ be an address such that $\heap(p) = (\cname'',\ftable)$.
Due to abstraction there may exist addresses $q \in \dom(heap)$ different from $p$ that alias with $p$.
Hence they are affected by the field update.
We introduce unsharing refinements (cf.~Definition~\ref{d:unsharing}) for all $q$, where $p \neq q \notin iu$.
%
%

\begin{definition}
\label{d:classinstance}
Let $\as=(\heap,\myframes,\isunshared)$ be a state and let $p$ be an address such that $\heap(p) = cn'$.
Let $cn \in subclasses(cn')$.
Furthermore, suppose $(\cname_1,id_1),\dots,(\cname_n,id_n)$
denote fields of $\cname$ (together with the defining classes). 
We perform the following \emph{class instance} steps, where the second takes care
of the case, where address $p$ is replaced by $\mynull$.
\begin{equation*}
  \infer{(\heap\{p \mapsto (\cname,ftable_1)\},\myframes,\isunshared)}{%
    (\heap,\myframes,\isunshared)%
  }
  \hspace{15mm}
  \infer{(\heap_2,\myframes_2,\isunshared)}{%
    (\heap,\myframes,\isunshared)%
  }
  \tpkt
\end{equation*}
Here $ftable_1((\cname_i,id_i)) \defsym v_i$ such that the type of the
abstract variable $v_i$ is defined in correspondence to the type
of field $(\cname_i, id_i)$, eg., a fresh $int$ variable for integer fields.
On the other hand we set $\heap_2$ ($\myframes_2$) equal to $\heap$ ($\myframes$),
but $p \notin \dom(\heap_2)$ and all occurrences of $p$ are replaced by $\mynull$.
\end{definition}

\begin{definition}
\label{d:unsharing}
Let $\as=(\heap,\myframes,\isunshared)$ and let $p$ and $q$ denote different addresses in $\heap$ such that $p \neq q \notin \isunshared$.
We perform the following \emph{unsharing} steps: 
The first case forces these addresses to be distinct.
The second case substitutes all occurrences of $q$ with $p$.
\begin{equation*}
    \infer{(\heap,\myframes,\isunshared \cup \{ \unshare{p}{q} \})}{%
      (\heap,\myframes,\isunshared)%
    }
    \hspace{15mm}
    \infer{(\heap',\myframes',\isunshared)}{%
      (\heap,\myframes,\isunshared)%
    }
    \tkom
\end{equation*}
where $\heap'$ ($\myframes'$) is equal to $\heap$ ($\myframes$) with all occurrences of $q$ replaced by $p$.
\end{definition}

\begin{figure}[ht]
\footnotesize
  \centering
  \begin{equation*}
  \begin{array}{lll}
    \mraisebox{\h\jload} & \infer{\absjvmstate{}{\heap}{\jvmframe{\loc(n) \cons \opstk}{\loc}{\cname}{\mname}{\PC+1}\cons \myframes}{\isunshared}}{\absjvmstate{}{\heap}{\jvmframe{\opstk}{\loc}{\cname}{\mname}{\PC}\cons \myframes}{\isunshared}}
    &
    \\[2mm]
    \mraisebox{\h\jiadd} & \infer{\absjvmstate{}{\heap}{\jvmframe{i_3 \cons \opstk}{\loc}{\cname}{\mname}{\PC+1}\cons \myframes}{\isunshared}}{\absjvmstate{}{\heap}{\jvmframe{i_2 \cons i_1 \cons \opstk}{\loc}{\cname}{\mname}{\PC}\cons \myframes}{\isunshared}}
    &
    \mraisebox{i_1 + i_2 = i_3}
    \\[2mm]
    \mraisebox{\h\jcmpeq} & \infer{\absjvmstate{}{\heap}{\jvmframe{b_3 \cons \opstk}{\loc}{\cname}{\mname}{\PC+1}\cons \myframes}{\isunshared}}{\absjvmstate{}{\heap}{\jvmframe{val_2 \cons val_1 \cons \opstk}{\loc}{\cname}{\mname}{\PC}\cons \myframes}{\isunshared}}
    &
    \mraisebox{(val_1 = val_2) \equiv b_3}
    \\[2mm]
   \mraisebox{\h\jiffalse~i} & \infer{\absjvmstate{}{\heap}{\jvmframe{\opstk}{\loc}{\cname}{\mname}{\PC+i}\cons \myframes}{\isunshared}}{\absjvmstate{}{\heap}{\jvmframe{\false \cons \opstk}{\loc}{\cname}{\mname}{\PC}\cons \myframes}{\isunshared}}
   &
   \\[2mm]
   & \infer{\absjvmstate{}{\heap}{\jvmframe{\opstk}{\loc}{\cname}{\mname}{\PC+1}\cons \myframes}{\isunshared}}{\absjvmstate{}{\heap}{\jvmframe{\true \cons \opstk}{\loc}{\cname}{\mname}{\PC}\cons \myframes}{\isunshared}}
   &
   \\[2mm]
   \mraisebox{\h{\jnew}~\cname'} & \infer{\absjvmstate{}{\heap'\{\addr \mapsto x\}}{\jvmframe{\addr \cons \opstk}{\loc}{\cname}{\mname}{\PC + 1} \cons \myframes}{\isunshared}}{\absjvmstate{}{\heap}{\jvmframez \cons \myframes}{\isunshared}}
   \\[2mm]
    \mraisebox{\h{\jputfield}~\fname~\cname'} & \infer{\absjvmstate{\varnothing}{\heap\{\addr \mapsto (\cname'', \ftable')\}}{\jvmframe{\opstk}{\loc}{\cname}{\mname}{\PC + 1} \cons \myframes}{\isunshared}}{\absjvmstate{\varnothing}{\heap}{\jvmframe{\val \cons \addr \cons \opstk}{\loc}{\cname}{\mname}{\PC} \cons \myframes}{\isunshared}}
    \\[2mm]
  \end{array}
  \end{equation*}
  \caption{Symbolic evaluations of Jinja bytecode instructions}
  \label{fig:symeval}
\end{figure}

\begin{example}
In Figure~\ref{fig:classinstance} we present an example detailing the need for
the given definition of class instantiation.
Here class \ttt{B} overrides method \ttt{m} inherited from class \ttt{A}.
We only know the static type of the parameter when analysing method \ttt{call(A a)}.
Method \ttt{call(A a)} accepts any instances of class \ttt{A} or any instances of a subclass of \ttt{A} as parameter.
In particular any instance of class \ttt{B}.
Due to the overridden method \ttt{call(A a)} does not terminate for instances of class \ttt{B}.
\end{example}
\begin{figure}[ht]
  \centering
  \begin{multicols}{2}
  \begin{lstlisting}
    class A{
      void m(){unit}
    }
    class B extends A{
      void m(){while(true)}
    }

    class C{
      void call(A a){a.m()}
      void main(){
        C c = new C();
        c.call(new B());
      }
    }; 
  \end{lstlisting}
  \end{multicols}
  \caption{All subclasses need to be considered.}
  \label{fig:classinstance}
\end{figure}

Let $\as, \as{'}$ and $\at$ be abstract states such that $\as{'}$ is obtained by zero or multiple refinement steps from $\as$.
Furthermore, suppose $\at$ is obtained from $\as{'}$ due to a symbolic evaluation.
Then we say $\at$ is obtained form $\as$ by an \emph{abstract computation}.

To prove correctness of an symbolic evaluation step, we have to show that $f^\ast(\gamma(\as)) \subseteq \gamma(\af(\as))$.
Hence,  it is enough to show that for all $s \in \gamma(\as)$ and $\JVMstep{s}{t}{}$ it follows that $t \in \gamma(\at)$, where $\at$ is obtained from a symbolic evaluation step, ie.,  $\at = \af(\as)$.
Similarly, to prove correctness of the refinement steps it is enough to show that for all $s \in \gamma(\as)$ there exists a state $\as_i$ obtained by a state refinement of $\as$ such that $s \in \gamma (\as_i)$.
Correctness of an abstract computation step follows from the correctness of refinement and symbolic evaluation steps.

\begin{lemma}
  \label{l:ref}
Let $\as \in \AStates$.
Suppose $\as_1, \ldots, \as_n$ is obtained by a state refinement from $\as$.
Then $\as \abstraction \as_i$ for all $\as_i$.
Furthermore, $s \in \gamma(\as)$ implies that there exists an abstract state $\as_i$ such that $s \in \gamma(\as_i)$.
\end{lemma}
\begin{proof}
  The claim follows easily by the definition of Boolean and class variables, and the fact that two addresses in the heap of $\as$ either alias or not.
\end{proof}

\begin{lemma}
\label{l:1}
Let $\as, \at \in \AStates$ such that $\at$ is obtained by a symbolic evaluation from $\as$.
Suppose $s \in \gamma(\as)$ and $\JVMstep{s}{t}{}$.
Then $t \in \gamma(\at)$.
\end{lemma}
\begin{proof}
The proof is straightforward in most cases; we only treat some informative ones.
Let $\as = (\h\heap, \h\myframe \cons \h\myframes, \isunshared)$ and
$s = (\heap, \myframe \cons \myframes)$.
By assumption the domain of $\h\myframe \cons \h\myframes$ and $\myframe \cons \myframes$ coincide.
\begin{itemize}
%
\item Consider $\h\jload~\num$. 
By assumption $\h{loc}(n) \absAux{m} loc(n)$.
In the abstract computation step $\h\loc(n)$ is loaded on to the top of the stack.
Obviously $\h\opstk_i(n) \absAux{m'} \opstk_i(n)$, where $\opstk_i$ represents the top of the stack.
Then $t \in \gamma(\at)$.
\item Consider $\h\jiadd$. Let $i_2,i_1$ denote the first two
stack elements of $\as$. 
Wlog.\@ suppose that $i_1$ is abstract.
By definition of the symbolic evaluation of $\h\jiadd$
we perform the step by introducing a new abstract integer $i_3$
and adding the constraint $i_3 = i_1 + i_2$. 
Then $t \in \gamma(\at)$, since $i_3 \absAux{} z$ for all numbers $z$.
\item Consider $\h\jiffalse~i$. Wlog.\@ let $\false$ be the 
top element of the stack of $\as$. Executing the symbolic step yields
a state $\at$, which is an abstraction of $t$ by assumption on $s$
and $\as$.
Then $t \in \gamma(\at)$.
%
%
\item Consider $\h\jputfield~\fname~\cname$ on address $p$.
  By assumption the instruction can be symbolically evaluated and $p$ does not alias with some address $q \in \dom(\h\heap)$ different from $p$.
  The only interesting case to consider is when $\h\heap(q)$ is a class variable and there exists 
  $s \in \gamma(\as)$ such that 
  $m(q) \reach[S] r \reachtr[S]  m(p)$, where $r \in \dom(\heap)$.
  Then $m(q)$ reaches $m(p)$ via $r$ and is affected by the update instruction.
  This does not matter, since $\h\heap(q)$ is also a class variable in $\at$, thus also representing the affected instance.
  Then $t \in \gamma(\at)$.
\item Consider $\h\jcmpeq$. 
  By assumption the instruction can be symbolically executed.
  That is the necessary refinement steps are already performed.
  Then $t \in \gamma(\at)$ follows directly.
\end{itemize}
\end{proof}

The next theorem is an immediate result of the lemma.
\begin{theorem}
\label{t:3}
Let $s$ and $t$ be JVM states, such that $\JVMexec{s}{t}$. 
Suppose $s \in \gamma(\as)$ for some state $\as$.
Then there exists an abstract computation of $\at$ from $\as$ such that $t \in \gamma(\at)$.
\end{theorem}

Theorem~\ref{t:3} formally proves the correctness of the proposed
abstract domain with respect to the operational semantics for Jinja,
established by Klein and Nipkow~\cite{KN06}.
In order to exploit this abstract domain we require a finite representation
of the abstract domain $\AStates$ induced by $\Program$. For that
we propose in the next section computation graphs as finite representations of all 
relevant states in $\AStates$, abstracting JVM states in $\Program$.

%% file: computationgraph.tex
\section{Computation Graphs}
\label{ComputationGraph}

In this section, we define \emph{computation graphs} as \emph{finite} representations
of the abstract domain $\AStates$ with respect to $\Program$.

\begin{definition}
\label{d:computationgraph}
A \emph{computation graph} $\CGraph=(\nodes{\CGraph},\edges{\CGraph})$ is a
directed graph with edge labels, where $\nodes{\CGraph} \subset \AStates$ 
and $\as \toss{\ell} \at \in \edges{\CGraph}$ if either 
$\at$ is obtained from $\as$ by an abstract computation or
$\as$ is an instance of $\at$. Furthermore, 
if there exists a constraint $C$ in the symbolic evaluation, then 
$\ell \defsym C$. For all other cases $\ell \defsym \varnothing$. 
We say that $\CGraph$ is the computation graph of program $\Program$ if
for all initial states $\start$ of $\Program$ there exists an abstract state
$\h{\start} \in \CGraph$ such that $\start \in \gamma(\h{\start})$.
\end{definition}

We obtain a finite representation of loops, if we suitably exploit the fact that any subset of $\AStates$ has a least upper bound.
The intution is best conveyed by an example.

\begin{example}
\label{ex:compgraph}
Consider the \ts{List} program from Example~\ref{ex:append} together
with the well-formed JBC program depicted in Figure~\ref{fig:listbc}. 
Figure~\ref{fig:compgraph} illustrates the computation graph of \ttt{append}.
For the sake of readability we omit the $val$ field of the list, the unsharing annotations and some intermediate nodes.

\begin{figure}[h!t]
  \centering
  \begin{tikzpicture}[]
    \scriptsize
    \node[state](A){
      \parbox{.43\textwidth}{
        $
        \begin{array}[ht]{l|l}
          \m{00} & \epsilon \mid \text{this} = o_1, \text{ys} = o_2, \text{cur} = \m{unit} \\
          & o_1 = \m{List(\text{List.next}} = o_3\m{)} \\
          I & o_2 = list, o_3 = list \\
        \end{array}
        $
      }
    };
    \node[state, below = .4cm of A](B){
      \parbox{.43\textwidth}{
        $
        \begin{array}[ht]{l|l}
          \m{04} & \epsilon \mid \text{this} = o_1, \text{ys} = o_2, \text{cur} = o_1 \\
          & o_1 = \m{List(\text{List.next}} = o_3 \m{)} \\
          A & o_2 = list, o_3 = list \\
        \end{array}
        $
      }
    };
    \node[state,right = .8cm of B,yshift=-1cm, dashed](B2){
      \parbox{.43\textwidth}{
        $
        \begin{array}[ht]{l|l}
          \m{04} & \epsilon \mid \text{this} = o_1, \text{ys} = o_2, \text{cur} = o_3  \\
          & o_1 = \m{List(\text{List.next}} = o_3 \m{)} \\
          & o_2 = list, o_4 = list \\
          B & o_3 = \m{List(\text{List.next}} = o_4 \m{)} \\
        \end{array}
        $
      }
    };
    \node[state, below = 0.4cm of B](B1){
      \parbox{.43\textwidth}{
        $
        \begin{array}[ht]{l|l}
          \m{04} & \epsilon \mid \text{this} = o_1, \text{ys} = o_2, \text{cur} = o_4 \\
          & o_1 = \m{List(\text{List.next}} = o_3 \m{)} \\
          & o_2 = list, o_3 = list, o_5 = list \\
          S & o_4 = \m{List(\text{List.next}} = o_5 \m{)} \\
        \end{array}
        $
      }
    };
    \node[state, below = 0.4cm of B2](D2){
      \parbox{.43\textwidth}{
        $
        \begin{array}[ht]{l|l}
          \m{04} & \epsilon \mid \text{this} = o_1, \text{ys} = o_2, \text{cur} = o_5  \\
          & o_1 = \m{List(\text{List.next}} = o_3 \m{)} \\
          & o_2 = list, o_3 = list, o_6 = list \\
          D & o_5 = \m{List(\text{List.next}} = o_6 \m{)} \\
        \end{array}
        $
      }
    };
    \node[state, below = 0.4cm of D2](C2){
      \parbox{.43\textwidth}{
        $
        \begin{array}[ht]{l|l}
          \m{07} & o_5, \m{null} \mid \text{this} = o_1, \text{ys} = o_2, \text{cur} = o_4  \\
          & o_1 = \m{List(\text{List.next}} = o_3 \m{)} \\
          & o_2 = list, o_3 = list, o_6 = list \\
          & o_4 = \m{List(\text{List.next}} = o_5 \m{)} \\
          C_1 & o_5 = \m{List(\text{List.next}} = o_6 \m{)}
        \end{array}
        $
      }
    };
    \node[state, below = 2.8cm of B1](C){
      \parbox{.43\textwidth}{
        $
        \begin{array}[ht]{l|l}
          \m{07} &  o_5, \m{null} \mid \text{this} = o_1, \text{ys} = o_2, \text{cur} = o_4  \\
          & o_1 = \m{List(\text{List.next}} = o_3 \m{)} \\
          & o_2 = list, o_3 = list, o_5 = list \\
          C & o_4 = \m{List(\text{List.next}} = o_5 \m{)}
        \end{array}
        $
      }
    };
    \node[state, below = .4cm of C](C1){
      \parbox{.43\textwidth}{
        $
        \begin{array}[ht]{l|l}
          \m{07} & \m{null}, \m{null} \mid \text{this} = o_1, \text{ys} = o_2, \text{cur} = o_4  \\
          & o_1 = \m{List(\text{List.next}} = o_3 \m{)} \\
          & o_2 = list, o_3 = list \\
          C_2 & o_4 = \m{List(\text{List.next}} =  \m{null)} \\
        \end{array}
        $
      }
    };
    \node[state, right = .8cm of C1](D1){
      \parbox{.43\textwidth}{
        $
        \begin{array}[ht]{l|l}
          \m{19} & o_4, o_2 \mid \text{this} = o_1, \text{ys} = o_2, \text{cur} = o_4  \\
          & o_1 = \m{List(\text{List.next}} = o_3 \m{)} \\
          & o_2 = list, o_3 = list \\
          E & o_4 = \m{List(\text{List.next}} =  \m{null)} \\
        \end{array}
        $
      }
    };
    \node[state, below = .4cm of C1](E1){
      \parbox{.43\textwidth}{
        $
        \begin{array}[ht]{l|l}
          \m{19} & o_4, o_2 \mid \text{this} = o_1, \text{ys} = o_2, \text{cur} = o_1  \\
          & o_1 = \m{List(\text{List.next}} =  \m{null)} \\
          E_1 & o_2 = list
        \end{array}
        $
      }
    };
    \node[state, below = .4cm of E1](F1){
      \parbox{.43\textwidth}{
        $
        \begin{array}[ht]{l|l}
          \m{--} & \epsilon \mid \text{this} = o_1, \text{ys} = o_2, \text{cur} = o_1  \\
          & o_1 = \m{List(\text{List.next}} = o_2 \m{)} \\
          F_1 & o_2 = list
        \end{array}
        $
      }
    };
    \node[state, right = .8cm of F1, yshift =-0.4cm](E2){
      \parbox{.43\textwidth}{
        $
        \begin{array}[ht]{l|l}
          \m{19} & o_4, o_2 \mid \text{this} = o_1, \text{ys} = o_2, \text{cur} = o_4  \\
          & o_1 = \m{List(\text{List.next}} = o_3 \m{)} \\
          & o_2 = list, o_3 = list\\
          E_3 & o_4 = \m{List(\text{List.next}} =  \m{null)}
        \end{array}
        $
      }
    };
    \node[state, below = .4cm of E2](F2){
      \parbox{.43\textwidth}{
        $
        \begin{array}[ht]{l|l}
          \m{--} & \epsilon \mid \text{this} = o_1, \text{ys} = o_2, \text{cur} = o_4  \\
          & o_1 = \m{List(\text{List.next}} = o_3 \m{)} \\
          & o_2 = list, o_3 = list \\
          F_3 & o_4 = \m{List(\text{List.next}} =  o_2 \m{)}
        \end{array}
        $
      }
    };
    \node[state, below = .4cm of F1](E3){
      \parbox{.43\textwidth}{
        $
        \begin{array}[ht]{l|l}
          \m{19} & o_4, o_2 \mid \text{this} = o_1, \text{ys} = o_2, \text{cur} = o_3  \\
          & o_1 = \m{List(\text{List.next}} = o_3 \m{)} \\
          & o_2 = list \\
          E_2 & o_3 = \m{List(\text{List.next}} =  \m{null)}
        \end{array}
        $
      }
    };
    \node[state, right = -0.3cm of E3, yshift=0.65cm, draw=none](E4){};
    \node[state, left = -1.3cm of D1, yshift=-0.6cm, draw=none](D3){};
    \node[state, below = .4cm of E3](F3){
      \parbox{.43\textwidth}{
        $
        \begin{array}[ht]{l|l}
          \m{--} & \epsilon \mid \text{this} = o_1, \text{ys} = o_2, \text{cur} = o_3  \\
          & o_1 = \m{List(\text{List.next}} = o_3 \m{)} \\
          & o_2 = list \\
          F_2 & o_3 = \m{List(\text{List.next}} =  o_2 \m{)}
        \end{array}
        $
      }
    };

    \path[->] (A) edge node {} (B);
    \path[->] (B) edge[right]  node {$\sqsubseteq$} (B1);
    \path[->, dashed] (B) edge[right]  (B2);
    \path[->, dashed] (B2) edge[below]  node {$\sqsupseteq$} (B1);
    \path[->] (B1) edge node {} (C);
    \path[->] (C) edge [] node {} (C1);
    \path[->] (C) edge [] node {} (C2);
    \path[->] (C1) edge node {} (D1);
    \path[->] (C2) edge node {} (D2);
    \path[->] (D2) edge[below]  node {$\sqsupseteq$} (B1);
    \path[->] (D1) edge node {} (E1);
    \path[->] (D1) edge node {} (E2);
    \path[->] (D3) edge node {} (E4);
    \path[->] (E1) edge node {} (F1);
    \path[->] (E2) edge node {} (F2);
    \path[->] (E3) edge node {} (F3);
  \end{tikzpicture}
  \caption{The (incomplete) computation graph of \ttt{append}.}
  \label{fig:compgraph}
\end{figure}

Consider the initial node $I$.
It is easy to see that $I$ is an abstraction of all concrete initial states, when $this$ is not $\mynull$.
We assume that $this$ is acyclic and initially do not share with $ys$.
Nodes $A$, $B$ and $S$ correspond to the situation described in Example~\ref{ex:A} and Example~\ref{ex:S}.
That is, node $A$ is obtained after assigning $cur$ to $this$ before any iteration of the loop,
node $B$ is obtained after exactly one iteration of the loop
and node $S = \lub\{A,B\}$.
Intermediate iterations are normally removed.
This is indicated by a dashed border for $B$.

After pushing the reference of $cur.next$ and $\mynull$ onto the operand stack, 
we reach node $C$.
At $\ttt{pc} = 7$ we want to compare the reference of $cur.next$ with $\mynull$.
But, $cur.next$ is not concrete.
Therefore, a class instance refinement is performed, yielding nodes $C_1$ and $C_2$.

First, we consider that $cur.next$ is not $\mynull$, but references an arbitrary instance, as illustrated in node $C_1$.
The step from $C_1$ to $D$ is trivial.
Let $id$ denote the identity function and $m = id(\nodes{S})$.
Then $m\{o_4 \mapsto o_5, o_5 \mapsto o_6\}$ is a morphism from $S$ to $D$.
Therefore, $D$ is an instance of $S$.
Second, we consider the case when $cur.next$ is $\mynull$, as depicted in node $C_2$.
Node $E$ is obtained from $C_2$ after loading registers $cur$ and $ys$ onto the stack.
At program counter $19$ a $\h\jputfield$ instruction is performed.
Therefore we perform a refinement according to Definition~\ref{d:unsharing}.
We obtain nodes $E_1, E_2$ and $E_3$.
In $E_1$, $this$ and $cur$ point to the same reference, in $E_2$ $this.next$ and $cur$ point to the same reference, and in $E_3$ the abstracted part from $cur$ is distinct from $this$, yet $this$ and $cur$ shares.
Nodes $F_1, F_2$ and $F_3$ are obtained after performing the $\h\jputfield$ instruction.
\end{example}

To concretise the employed strategy, note that whenever
we are about to finish a loop, we attempt to use an instance refinement
to the state starting this loop. If this fails, for example in an attempted step from 
$B$ to $A$ in Example~\ref{ex:compgraph}, we widen the corresponding state. Here we
collect all states that need to be abstracted and join them to obtain
an abstraction.
Complementing the proposed strategy, we restrict the applications of refinements, such that refinement steps are only performed if no other steps are applicable.
We say that this strategy is an \emph{eager strategy}.
The next lemma shows that if an eager strategy is followed we are guaranteed to obtain a \emph{finite} computation graph.

\begin{lemma}
\label{l:finite}
Let $\CGraph$ be the computation graph of a program $P$ such
that in the construction of $\CGraph$ an eager strategy is applied.
Then $\CGraph$ is finite.  
\end{lemma}
\begin{proof}
We argue indirectly. Suppose the computation
graph $\CGraph$ of $P$ is infinite. This is only possible if
there exists an initial state $\start$ of $\Program$ that is non-terminating, which
implies that starting from $\start$ we reach a loop in $\Program$ that is
called infinitely often. As $\CGraph$ is infinite this implies that the
widening operation for this loop gives rise to an infinite sequence of
states $(\as_j)_{j \geqslant 0}$ such that $\as_j \properinstance \as_{j+1}$
for all $j$. However, this is impossible as any ascending chain of
abstract states eventually stabilises, cf.~Lemma~\ref{l:chain}. 
\end{proof}

Let $\CGraph$ be a computation graph. We write $\GCstep{\as}{\at}$ to indicate that state 
$\at$ is directly reachable in $G$ from $\as$. 
Sometimes we want to distinguish whether $\at$ is obtained 
by a refinement (denoted as $\GCstepref{\as}{\at}$) or
by a symbolic evaluation (denoted as $\GCstepeva{\as}{\at}$),
or whether $\as$ is an instance of $\at$ (denoted as $\GCstepins{\as}{\at}$). 
If $\at$ is reachable from $\as$ in $\CGraph$
we write $\GCexec{\as}{\at}$. If $\as \not= \at$ this is denoted by $\GCexecir{\as}{\at}$.

\begin{lemma}
\label{l:11}
Let $s, t \in \States$ such that $\JVMstep{s}{t}$.
Let $G$ denote the computation graph of $P$, and $\as, \at \in G$.
Suppose $s \in \gamma(\as)$, then there exists $\at$ such that $t \in \gamma(\at)$ and ${\as} \reachtr[\text{ins}] \cdot \reachtr[\text{ref}] \cdot \reach[\text{eva}] \at$. 
\end{lemma}
\begin{proof}
  By construction of $G$ we have to consider two cases:
  Suppose $\at$ is obtained by an abstract computation from $\as$.
  We employ Lemma~\ref{l:1} to conclude that $t \in \gamma(\at)$.
  Then $\as \reachtr[\text{ref}] \cdot \reach[\text{eva}] \at$.
  Next, suppose $\at$ is obtained by an abstract computation from $\as{'}$, where $\as \instance \as{'}$.
  Hence, we also have $s \in \gamma(\as{'})$.
  We employ Lemma~\ref{l:1} to conclude that $t \in \gamma(\at)$.
  Then $\as \reachtr[\text{ins}] \cdot \reachtr[\text{ref}] \cdot \reach[\text{eva}] \at$.
  Since $G$ is finite we conclude that ${\as} \reachtr[\text{ins}] \cdot \reachtr[\text{ref}] \cdot \reach[\text{eva}] \at$ has finitely many instance and refinement steps, only depending on $G$.
\end{proof}

We arrive at the main result of this section. 
\begin{theorem}
\label{t:1}
Let $\start, t \in \States$ and suppose $\JVMexec{\start}{t}$, where
the runtime of the execution is $m$. 
Let $\CGraph$ denote the computation graph of $\Program$ obtained from some initial state 
$\astart$ such that $\start \in \gamma(\astart)$.
Then there exists an abstraction $\at \in G$ and a path ${\astart} \reachtr[G] {\at}$ of length $m'$ such that $m \leqslant m' \leqslant K \cdot m$.
Here constant $K \in \N$ only depends on $\CGraph$.
\end{theorem}
\begin{proof}
By induction on $m$ (employing Lemma~\ref{l:11}), 
we conclude the existence of state $\at$ such that 
$\GCexec{\start}{\at}$. 
Hence, the first part of the theorem follows.
Furthermore by Lemma~\ref{l:11} there exists $m'$ such that $m \leqslant m' \leqslant K \cdot m$.
\end{proof}

%% file: rewriting.tex
\section{Constrained Rewrite Systems}
\label{CTRS}

Let $\CGraph$ be the computation graph for program $\Program$
with initial state $\ai$; $\CGraph$ is kept fixed for the
remainder of the section.
In the following we describe the translation from $\CGraph$ into
a \emph{constrained term rewrite system} (\emph{cTRS} for short). Our
definition is a variation of cTRSs as for example defined by Falke and
Kapur~\cite{FK09,FKS11} or Sakata et al.~\cite{SNS11}. 
Recently, Kop and Nishida introduced a very general formalism of term rewrite systems with constraints, termed \emph{logical constrained term rewrite systems (LCTRSs)}~\cite{KN13}.
The proposed notion of cTRSs is not directly interchangeable with LCTRSs, yet the rewrite system resulting from the transformation could also be formalised as LCTRS.
The here proposed transformation is inspired by~\cite{OBEG10}. Otto et al.~transform 
\emph{termination graphs} into \emph{integer term rewrite systems} (\emph{ITRSs} for short)~\cite{FGPSF09}. 

Let $\GS$ be a (not necessarily finite) sorted signature, let $\VS'$ 
denote a countably infinite set of sorted variables. 
Furthermore let $\theory$ denote a theory over $\GS$. 
Quantifier-free formulas over $\GS$ are called \emph{constraints}. 
Suppose $\FS$ is a sorted signature that extends $\GS$ and let 
$\VS \supseteq \VS'$ denote an extension of the variables in $\VS'$.
Let $\TA(\FS,\VS)$ denote the set of \emph{(sorted) terms} over the 
signature $\FS$ and $\VS$. Note that the sorted signature is necessary
to distinguish between \emph{theory} variables that are to be interpreted over 
the theory $\theory$ and \emph{term} variables whose interpretation is free.
A \emph{constrained rewrite rule}, denoted as $\constraint{l}{r}{C}$, 
is a triple consisting of terms $l$ and $r$, together with a constraint $C$.
We assert that $l \not\in \VS$, but do \emph{not} require that $\Var(l) \supseteq
\Var(r) \cup \Var(C)$, where $\Var(t)$ ($\Var(C)$) denotes the variables
occurring in the term $t$ (constraint $C$). A \emph{constrained term rewrite
system} (\emph{cTRS}) is a finite set of constrained rewrite rules.

Let $\RS$ denote a cTRS. A context $D$ is a term with exactly one occurrence 
of a \emph{hole} $\hole$, and $D[t]$ denotes the term obtained by replacing 
the hole $\hole$ in $D$ by the term $t$.
A substitution $\sigma$ is a function that maps 
variables to terms, and $t\sigma$ denotes
the homomorphic extension of this function to terms.
We define the rewrite relation $\rsrew$ as follows.
For terms $s$ and $t$, $s \rsrew t$ holds, if there exists a context
$D$, a substitution $\sigma$ and a constrained rule $\constraint{l}{r}{C} \in \RS$
such that $s \theoryunify D[l\sigma]$ and $t=D[r\sigma]$ with 
$\theory \proves C\sigma$. Here $\theoryunify$ denotes unification modulo
$\theory$.
For extra variables $x$, possibly occurring in $t$, we demand that $\sigma(x)$ is in normal-form.

We often drop the reference to the cTRS $\RS$, if no confusion can arise
from this. A function symbol in $\FS$ is called \emph{defined} if $f$ occurs
as the root symbol of $l$, where $\constraint{l}{r}{C} \in \RS$. Function
symbols in $\FS \setminus \GS$ that are not defined, are called \emph{constructor}
symbols, and the symbols in $\GS$ are called \emph{theory} symbols.

A cTRS $\RS$ is called \emph{terminating}, if the relation $\rsrew$ is well-founded.
For a terminating cTRS $\RS$, we define its \emph{runtime complexity}, denoted
as $\rctrs$. We adapt the runtime complexity with respect to a standard
TRS suitable for cTRS $\RS$. (See~\cite{HM08} for
the standard definition.)
The \emph{derivation height} of a term $t$ (with respect to $\RS$)
is defined as the maximal length of a derivation (with respect to $\RS$) 
starting in $t$. The derivation height of $t$ is denoted as $\dheight(t)$.
Note that $\rsrew$ is not necessarily finitely branching for finite cTRSs, as fresh variables on the right-hand side of a rule can occur.

\begin{definition}
We define the \emph{runtime complexity} (with respect to $\RS$) as follows:
\begin{equation*}
  \rctrs(n) \eqk \max \{ \dheight(t) \mid \text{$t$ is basic and 
    $\size{t} \leqslant n$}\}
\tkom
\end{equation*}
where a term $t = f(t_1,\dots,t_k)$ is called \emph{basic}
if $f$ is defined, and the terms $t_i$ are only built over 
constructor, theory symbols, and variables.
We fix the size measure $\size{\cdot}$ below.
\end{definition}

In the following we are only interested in cTRS over a specific
theory $\theory$, namely Presburger arithmetic, that is, we
have $\theory \proves C$, if all ground instances of the 
constraint $C$ are valid in Presburger arithmetic. 
Recall, that Presburger arithmetic is decidable. 
If $\theory \proves C$, then
$C$ is \emph{valid}. On the other hand, if there exists a substitution
$\sigma$, such that $\theory \proves C\sigma$, then $C$ is \emph{satisfiable}. 

To represent the basic operations in 
the Jinja bytecode instruction set (cf.~Figure~\ref{fig:jbc})
we collect the following connectives and truth constants in
$\GS$: $\land$, $\lor$, $\lnot$, $\true$, and $\false$, together
with the following relations and operations: $=$, $\not=$, $\geqslant$, $+$, $-$. 
Furthermore, we add infinitely many constants to represent integers.
We often write $l \to r$ instead of 
$\constraint{l}{r}{\true}$. As expected $\GS$ makes use of two
sorts: $\m{bool}$ and $\m{int}$.
We suppose that all abstract variables $X_1,X_2,\dots$
are present in the set of variables $\VS$, where abstract integer (Boolean)
variables are assigned sort $\m{int}$ ($\m{bool}$) and all other
variables are assigned sort $\m{univ}$. 
The remaining elements of the signature $\FS$ will be defined in the
course of this section. As the signature of these function
symbols is easily read off from the translation given below, in the
following the sort information is left implicit, to simplify the presentation.

The size of a term $t$, denoted as $\size{t}$ is defined as follows:
\begin{equation*}
  \size{t} \defsym
  \begin{cases}
    1 & \text{if $t$ is a variable}\\
    \abs(t) & \text{if $t$ is an integer}\\
    1 + \sum_{i=1}^n \size{t_i} & \text{if $t=f(t_1,\dots,t_n)$ and $f$ is not
    an integer}
    \tpkt
  \end{cases}
\end{equation*}

In the next definition, we show how a state becomes representable
as term over~$\FS$.

\begin{definition}
Let $\as=(\heap,\myframes,\isunshared)$ be a state
and let the index sets $\stkfamily$ and 
$\locfamily$ be defined as above.
Suppose $v$ is a value.
Then the value $v$ is translated as follows:
  \begin{align*}
    \tval(v) & \defsym 
    \begin{cases}
      \mynull  & \text{if $v \in \{\unit,\mynull\}$}
      \\
      v        & \text{if $v$ is a non-address value, except $\unit$ or
      $\mynull$} 
      \\
      \tobj(v) & \text{if $v$ is an address}
      \tpkt
    \end{cases}
    \\
\intertext{Let $a$ be an address. Then $a$ is translated as follows:}
  \tobj(a) &\defsym 
  \begin{cases}
    x &
    \begin{minipage}[t]{40ex}
      if $a$ is maybe-cyclic and
      $x$ is a fresh variable
    \end{minipage}
    \\[3mm]
    x   & 
    \begin{minipage}[t]{40ex}
      if $\heap(a)$ denotes an abstract variable~$x$
    \end{minipage}
    \\
    \cname(\tval(v_1), \ldots, \tval(v_n)) & \text{if $\heap(a) = (\cname, \ftable)$}
    \tpkt
  \end{cases}
\end{align*}
Here we suppose in the last case that 
$\dom(\ftable) = \{(\cname_1,id_1),\ldots, (\cname_n,id_n)\}$ and 
for all $1 \leqslant i \leqslant n$: $\ftable((\cname_i,id_i) = v_i$.
Finally, to translate the state $s$ into a term, it suffices to translate
the values of the registers and the operand stacks of all frames in the
list $\myframes$. 
Let $(\opstk, i,j) \in \stkfamily$ such that
$\opstk_i(j)$ denotes the $j^{\text{th}}$ value in the operation
stack of the $i^{\text{th}}$ frame in $\myframes$. Similarly for 
$(\loc, i',j') \in \locfamily$. Then we set
\begin{equation*}
  \tst(s) \defsym [\tval(\opstk_1(1)),\dots,\tval(\opstk_k(\card{\opstk_k}))),
  \tval(\loc_1(1)),\dots,\tval(\loc_k(\card{\loc_k}))] \tkom
\end{equation*}
where the list $[\dots]$, is formalised by an auxiliary binary symbol
$\cons$ and the constant $\nil$.
\end{definition}

\begin{example}
Consider the simplified presentation of state $C$ in Figure~\ref{fig:compgraph}.
Then $\tst(C)$ yields following term:
\begin{equation*}
  \tst(C) = [list5, \mynull,\m{List}(list3), list2, \m{List}(\m{List}(list5))] \tpkt
\end{equation*}
\end{example}

Note that we can omit the information of the defining classes of the fields,
since this is already captured in the symbolic evaluation. Furthermore,
observe that our term representation can only fully represent acyclic data.
In this sense, the term representation of a state $s$
is less general, than its graph-based representation. However, we still
obtain the following lemma.

\begin{lemma}
\label{l:9}
Let $\as$ and $\at$ be abstract states. If $\at \instance \as$, then there
exists a substitution $\sigma$ such that $\tst(\at) = \tst(\as)\sigma$.
\end{lemma}
\begin{proof}
Let $\aS$ and $\aT$ be the state graphs of $\as$ and $\at$, respectively. 
By assumption there exists a morphism $m \colon \aS \to \aT$.
The lemma is a direct consequence of the following observations:
\begin{itemize}
\item Consider the terms $\tst(\as)$ and $\tst(\at)$. By definition these terms
encode the standard term representations of the graphs $\aS$ and $\aT$.
\item Let $u$ and $v$ be nodes in $\aS$ and $\aT$ such that
$m(u) = v$. The label of $u$ (in $\aS$) can only be distinct from
the label of $v$ (in $\aT$), if $\lab{\aS}(u)$ is an abstract variable or
$\mynull$. In the former case $\tval(\lab{\aS}(u))$ is again a variable
and the latter case implies that $\lab{\aT}(v) = \unit$. Thus in both
cases, $\tval(\lab{\aS}(u))$ matches $\tval(\lab{\aT}(v))$.
\item By correctness of our abstraction, we have $m(u)$ is maybe-cyclic, if $v$ is maybe-cyclic.
In this case $\tval(\lab{\aS}(u))$ and $\tval(\lab{\aT}(v))$ are fresh variables.
Hence, $\tval(\lab{\aS}(u))$ matches $\tval(\lab{\aT}(v))$.
\end{itemize}
\end{proof}

The next lemma relates the size of a state 
to its term representation and vice versa. 

\begin{lemma}
\label{l:8}
Let $s=(\heap,\myframes)$ be a state such that $\heap$ does not admit cyclic data structures.
Then $\size{\tst(\beta(s))} = \statesize{s}$.
\end{lemma}
\begin{proof}
As a consequence of Definition~\ref{d:size}
and the above proposed variant of the term complexity 
we see that $\size{\tst(\beta(s))} = \statesize{s}$ 
for all states $s$.
\end{proof}

\begin{lemma}
\label{l:8b}
Let $s=(\heap,\myframes)$ be a state such that $\heap$ may contain cyclic data structures. 
Then $\size{\tst(\beta(s))} \leqslant \statesize{s}$ and therefore $\size{\tst(\beta(s))} \in \bO(\statesize{s})$.
\end{lemma}
\begin{proof}
Follows from the previous lemma and the fact that addresses bounded to cyclic data structures are replaced by fresh variables.
\end{proof}

Let $\CGraph$ be a computation graph.
For any state $\as$ in $\CGraph$ we introduce a new function symbol $\funsym{\as}$.
Suppose $\tst(\as) = [\as_1,\dots,\as_n]$. To ease presentation we write
$\funsym{\as}(\tst(\as))$ instead of $\funsym{\as}(\as_1,\dots,\as_n)$.
\begin{definition}
\label{d:trs}
Let $\CGraph$ be a finite computation graph and $\as = (\heap, \myframes, \isunshared)$ and $\at$ be states in $\CGraph$.
We define the constrained rule \emph{corresponding} to the edge $(\as,\at)$, denoted by $\corrrule(\as,\at)$, as follows:
\begin{equation*}
\corrrule(\as,\at) = 
  \begin{cases}
    \funsym{\as}(\tst(\as)) \to \funsym{\at}(\tst(\as)) & \text{if $\as \instance \at$}\\
    \funsym{\as}(\tst(\at)) \to \funsym{\at}(\tst(\at)) & \text{if $\at$ is a state refinement of $\as$} \\
    \constraint{\funsym{\as}(\tst(\as))}{\funsym{\at}(\tst(\at))}{\tval(C)} & \text{the edge is labelled by $C$} \\
    \funsym{\as}(\tst(\as)) \to \funsym{\at}(\tst^\ast(\at)) & 
    \begin{minipage}[t]{30ex}
      $\as$ corresponds to a $\h\jputfield$ on address $p$, $\heap(q)$ is variable $cn$, and $q$ may-reach $p$
    \end{minipage}
    \\[2mm]
    \funsym{\as}(\tst(\as)) \to \funsym{t}(\tst(\at)) & 
     \text{otherwise} \tpkt
  \end{cases}
\end{equation*}
Here $\tval(C)$ denotes the standard extension of the mapping $\tval$ to labels
of edges and $\tst^\ast$ is defined as $\tst$ but employs fresh variables for any 
reference $q$ that may-reach the object that is updated.
The cTRS obtained from $\CGraph$ consists of rules $\corrrule(\as,\at)$ for all edges $\as \to \at \in \CGraph$.
\end{definition}

\begin{example}
  Figure~\ref{fig:appendtrs} illustrates the cTRS obtained from the computation graph of Example~\ref{ex:compgraph}.
We use following conventions: $\m{L}$ denotes the list constructor symbol and $l$ followed by a number a list variable.
In the last rule $l4$ is fresh on the right-hand side.
This is because we update $cur$ and have a side-effect on $this$ that is not directly observable in the abstraction.
\begin{figure}[ht]%
{ %
\renewcommand*{\mynull}{\m{null}}
\newcommand*{\iter}{\m{Iter}}
\newcommand*{\listx}[1]{\m{L}(\ensuremath{#1)}}
\newcommand*{\listinstance}[1]{\m{L}(\ensuremath{\listvar{#1})}}
\newcommand*{\listinstancenull}{\m{L}(\ensuremath{\mynull})}
\newcommand*{\listdinstance}[1]{\m{L}(\m{L}(\ensuremath{\listvar{#1}}))}
\newcommand*{\listvar}[1]{\ensuremath{l{#1}}}
\begin{align*}
  \funsym{I}(\listinstance{3}, \listvar{2}, \mynull) & \to \funsym{A}(\listinstance{3}, \listvar{2}, \listinstance{3}) \\
  \funsym{A}(\listinstance{3}, \listvar{2}, \listinstance{3}) & \to \funsym{S}(\listinstance{3}, \listvar{2}, \listinstance{3}) \\
  \funsym{S}(\listinstance{3}, \listvar{2}, \listinstance{5}) & \to \funsym{C}(\listvar{5},\mynull, \listinstance{3}, \listvar{2}, \listinstance{5}) \\
  \funsym{C}(\listinstance{6}, \mynull, \listinstance{3}, \listvar{2}, \listdinstance{6}) & \to \funsym{C_1}(\listinstance{6}, \mynull, \listinstance{3}, \listvar{2}, \listdinstance{6})\\
  \funsym{C}(\mynull, \mynull, \listinstance{3}, \listvar{2}, \listinstancenull) & \to \funsym{C_2}(\mynull, \mynull, \listinstance{3}, \listvar{2}, \listinstancenull)\\
  \funsym{C_1}(\listinstance{6}, \mynull, \listinstance{3}, \listvar{2}, \listdinstance{6}) & \to \funsym{D}(\listinstance{3}, \listvar{2}, \listinstance{6})\\
  \funsym{D}(\listinstance{3}, \listvar{2}, \listinstance{6}) & \to \funsym{S}(\listinstance{3}, \listvar{2}, \listinstance{6})\\
  \funsym{C_2}(\mynull, \mynull, \listinstance{3}, \listvar{2}, \listinstancenull) & \to \funsym{E}(\listinstancenull, \listvar{2}, \listinstance{3}, \listvar{2}, \listinstancenull)\\
  \funsym{E}(\listinstancenull, \listvar{2}, \listinstancenull, \listvar{2}, \listinstancenull) & \to \funsym{E_1}(\listinstancenull, \listvar{2}, \listinstancenull, \listvar{2}, \listinstancenull) \\
  \funsym{E_1}(\listinstancenull, \listvar{2}, \listinstancenull, \listvar{2}, \listinstancenull) & \to \funsym{F_1}(\listinstance{2}, \listvar{2}, \listinstance{2}) \\
  \funsym{E}(\listinstancenull, \listvar{2}, \listx{\listinstancenull}, \listvar{2}, \listinstancenull) & \to \funsym{E_2}(\listinstancenull, \listvar{2}, \listx{\listinstancenull}, \listvar{2}, \listinstancenull) \\
  \funsym{E_2}(\listinstancenull, \listvar{2}, \listx{\listinstancenull}, \listvar{2}, \listinstancenull) & \to \funsym{F_2}(\listdinstance{2}, \listvar{2}, \listinstance{2}) \\
  \funsym{E}(\listinstancenull, \listvar{2}, \listinstance{3}, \listvar{2}, \listinstancenull) & \to \funsym{E_3}(\listinstancenull, \listvar{2}, \listinstance{3}, \listvar{2}, \listinstancenull) \\
  \funsym{E_3}(\listinstancenull, \listvar{2}, \listinstance{3}, \listvar{2}, \listinstancenull) & \to \funsym{F_3}(\listinstance{4}, \listvar{2}, \listinstance{2})
\end{align*}
}%
\caption{The cTRS of \ttt{append}.}
\label{fig:appendtrs}
\end{figure}
\end{example}

In the following we show that the rewrite relation of the obtained cTRS safely approximates the concrete semantics of the concrete domain.
We first argue informally:
\begin{itemize}
  \item 
By Lemma~\ref{l:11} there exists a path ${\as} \reachtr[\text{ins}] \cdot \reachtr[\text{ref}] \cdot \reach[\text{eva}] \at$ in $G$ for $\JVMstep{s}{t}$ such that $s \in \gamma(\as)$ and $t \in \gamma(\at)$.
\item 
  Together with Lemma~\ref{l:9} we have to show that $\funsym{\as}(\tst(\beta(s))) \rstrew \funsym{\at}(\tst(\beta(t)))$.
\item We do this by inspecting the rules obtained from the transformation.
  We will see that instance steps and refinement steps do not modify the term instance.
  In case of evaluation steps the effect is either directly observable in the abstract state, as it happens for $\h\jpush$ for example, or indirectly by requiring that the substitution is conform with the constraint.
  In the case of the $\h\jputfield$ instructions we have to find a suitable substitution for fresh variables to accommodate possible side-effects.
\end{itemize}

\begin{lemma}
\label{l:7}
Let $\as$ and $\at$ be states in $\CGraph$
connected by an edge $\as \toss {\ell} \at$ from $\as$ to $\at$. Suppose $s \in \States$
with $s \in \gamma(\as)$. Suppose further that if the constraint $\ell$ labelling
the edge is non-empty, then $s$ satisfies $\ell$. Moreover, if $\as \toss {\ell} \at$
follows due to a refinement step, then $s$ is consistent with the chosen
refinement.
Then there exists $t \in \gamma(\at)$ such that
$\funsym{\as}(\tst(s')) \rsrew[\corrrule(\as,\at)] \funsym{\at}(\tst(t'))$ with $s' = \beta(s)$, $t' = \beta(t)$.
\end{lemma}
\begin{proof}
The proof proceeds by case analysis on the edge $\as \toss{\ell} \at$ in $\CGraph$,
where we only need to consider the following four cases. The argument for the 
omitted fifth case is very similar to the third case.
\begin{itemize}
\item \emph{Case} $\as \toss{\ell} \at$, as $\as \instance \at$; $\ell = \varnothing$.
By assumption $s' \instance \as \instance \at$. Hence, $s \in \gamma(\at)$ by
transitivity of the instance relation. By Lemma~\ref{l:9} there
exists a substitution $\sigma$ such that $\tst(s') = \tst(\as)\sigma$. In sum,
we obtain:
\begin{equation*}
  \funsym{\as}(\tst(s')) = \funsym{\as}(\tst(\as))\sigma \rsrew[\corrrule(\as,\at)] 
  \funsym{\at}(\tst(\as))\sigma = \funsym{\at}(\tst(t')) \tkom
\end{equation*}
where we set $t' \defsym s'$. 

\item \emph{Case} $\as \toss{\ell} \at$, as $\at$ is a refinement of $\as$; $\ell = \varnothing$.
By assumption $s' \instance \as$ and $s$ is concrete. 
Hence, $s' \instance \at$ 
by definition of $\at$. Again by Lemma~\ref{l:9} there exists a substitution $\sigma$, 
such that $\tst(s') = \tst(\at)\sigma$. In sum, we obtain:
\begin{equation*}
  \funsym{\as}(\tst(s')) = \funsym{\as}(\tst(\at))\sigma \rsrew[\corrrule(\as,\at)] 
  \funsym{\at}(\tst(\at))\sigma = \funsym{\at}(\tst(t')) \tkom
\end{equation*}
where we again set $t' \defsym s'$. 

\item \emph{Case} $\as \toss{\ell} \at$, as $\at$ is the result of the symbolic
  evaluation of $\as$ and $\ell = C \not= \varnothing$.
By assumption $s$ satisfies the constraint $C$. More precisely, there
exists a substitution $\sigma$ such that $\tst(s') = \tst(\as)\sigma$ and
$\theory \proves \tval(C)\sigma$. We obtain:
\begin{equation*}
  \funsym{\as}(\tst(s')) = \funsym{\as}(\tst(\as))\sigma \rsrew[\corrrule(\as,\at)] 
  \funsym{\at}(\tst(\at))\sigma \tpkt
\end{equation*}
Let $t$ be defined such that $\JVMstep{s}{t}$. By Lemma~\ref{l:1} we
obtain $t' \instance \at$ and by inspection of the proof of Lemma~\ref{l:1}
we observe that $\tst(t') = \tst(\at)\sigma$. In sum, $\funsym{\as}(\tst(s')) \rsrew[\corrrule(\as,\at)] \funsym{\at}(\tst(t'))$.

\item \emph{Case} $\as \toss{\ell} \at$, as $\at$ is the result of a $\h\jputfield$ instruction
on $p$ and there exists an address $q$ in $\as$ that may-reaches $p$.
By assumption $s' \instance \as$ and thus $\tst(s') = \tst(\as)\sigma$ for
some substitution $\sigma$. Let $t$ be defined such that $\JVMstep{s}{t}$.
Due to Lemma~\ref{l:1}, we have $t' \instance \at$ and thus there
exists a substitution $\tau$ such that $\tst(t') = \tst^\ast(\at)\tau$. 

Consider the rule $\funsym{\as}(\tst(\as)) \to \funsym{\at}(\tst^\ast(\at))$. 
By definition address $q$ points in $\as$ to an abstract variable $x$
such that $x$ occurs in $\tst(\as)$ and $\tst(\at)$. Furthermore, $x$ is 
replaced by an extra variable $x'$ in $\tst^\ast(\at)$. Wlog., 
we assume that $x'$ is the only extra variable in $\tst^\ast(\at)$.
Let $m$ be a morphism such that $m \colon \as \to s'$
and $m(q) \reachtir[] m(p)$. By definition of
$\h\jputfield$, $m(p)$ and $m(q)$ exist in $t'$ and only the part
of the heap reachable from these addresses can differ in $s'$ and $t'$.

In order to show the admissibility of the rewrite step 
$\funsym{\as}(\tst(s')) \to \funsym{\at}(\tst(t'))$ we define a substitution
$\rho$ such that $\tst(\as)\rho = \tst(s')$ and $\tst^\ast(\at)\rho = \tst(t')$.
We set:
\begin{equation*}
  \rho(y) \defsym
  \begin{cases}
    \tau(x) & \text{if $y = x'$}\\
    \sigma(y) & \text{otherwise}
    \tpkt
  \end{cases}
\end{equation*}
Then $\tst(\as)\rho = \tst(s')$ by definition as $x' \not\in \Var(\as)$.
On the other hand $\tst^\ast(\at)\rho = \tst(t')$ follows as the definition
of $\rho$ forces the correct instantiation of $x'$ and Lemma~\ref{l:1}
in conjunction with Lemma~\ref{l:9}
implies that $\sigma$ and $\tau$ coincide on the portion of the
heap that is not changed by the field update.

\end{itemize}
\end{proof}

The next lemma emphasises that any execution step is represented by finitely many but at least one rewrite steps in $\RS$.
\begin{lemma}
\label{l:10}
Let $\as \in \CGraph$ and $s \in \States$ such that $s \in \gamma(\as)$.
Then $\JVMstep{s}{t}$ implies that there exists
a state $\at \in \CGraph$ such that $t \in \gamma(\at)$ and 
$\funsym{\as}(\tst(\beta(s))) \rsnrew{{} \leqslant K} \funsym{\at}(\tst(\beta(t)))$.
Here $K$ depends only on $\CGraph$ and $\rsnrew{{} \leqslant K}$ denotes at least one and at most $K$ many rewrite steps in~$\RS$.
\end{lemma}
\begin{proof}
The lemma follows from the proof of Lemma~\ref{l:11} and Lemma~\ref{l:7}.  
\end{proof}

We arrive at the main result of this thesis.
\begin{theorem}
\label{t:2}
Let $s, t \in \States$. Suppose $\JVMexec{s}{t}$, where $s$
is reachable in $\Program$ from some initial state $\start$. 
Set $s' = \beta(s)$, $t' = \beta(t)$.
Then there exists $\as, \at \in \AStates$ and a derivation
$\funsym{\as}(\tst(s')) \rstrew \funsym{\at}(\tst(t'))$ 
such that $s \in \gamma(\as)$ and $t \in \gamma(\at)$. 
Furthermore, for all $n$: $\rcjvm(n) \in \bO(\rctrs(n))$.  
\end{theorem}
\begin{proof}
The existence of $\as$ follows from the correctness of abstract computation together with the construction of the computation graph.
Let $m$ denote the runtime of the execution 
$\JVMexec{s}{t}$. Then by induction on $m$ in conjunction with Lemma~\ref{l:10} 
we obtain the existence of a state $\at$ such that $t' \instance \at$
and a derivation:
\begin{equation}
  \label{eq:1}
  \funsym{\as}(\tst(s')) \rsnrew{{} \leqslant K \cdot m} \funsym{\at}(\tst(t')) 
  \tpkt
\end{equation}
Here the constant $K$ depends only on $\CGraph$.
In particular we have $\funsym{s}(\tst(s')) \rstrew \funsym{t}(\tst(t'))$
from which we conclude the first part of the theorem.

To conclude the second part, let $n$ be arbitrary and
suppose $m$ denotes the runtime of the execution $\JVMexec{\start}{t}$,
where $\statesize{\start} \leqslant n$. 
We set $\start' = \beta(\start)$.
As $\CGraph$ is the computation graph of $\Program$ we obtain 
$\start' \instance \astart$. 
From Lemma~\ref{l:8b} it follows that $\size{\tst(\beta(\start))} \leqslant \statesize{\start}$.
Specialising~\eqref{eq:1} to $\astart$ and $\start'$ yields 
$\funsym{\astart}(\tst(\start')) \rsnrew{{} \leqslant K \cdot m} \funsym{\at}(\tst(t'))$.
Thus we obtain 
\begin{equation*}
  \rcjvm(\statesize{i}) = m \leqslant K \cdot m \leqslant  \rctrs(\size{\tst{(\beta(\start))}}) \leqslant \rctrs(\statesize{i}) 
  \tpkt
\end{equation*}
%
%
%
\end{proof}

It is tempting to think that the precise bound on the number of
rewrite steps presented in Lemma~\ref{l:10} should translate to
a linear simulation between JVM executions and rewrite derivation.
Unfortunately this is not the case as the transformation is 
not termination preserving.
For this consider Figure~\ref{fig:notterm}.
\begin{figure}[h!t!]
\center{
\begin{minipage}[t]{0.5\textwidth}
\begin{lstlisting}[firstnumber=1]       
  class List{ List next; }

  class Main{
    void inits(List ys){
      while(ys.next != null){
        List cur = ys;
          while(cur.next.next != null){
            cur = cur.next
          }
        cur.next = null;
        }
    }
  }
\end{lstlisting}
\end{minipage}
}
\caption{The \ttt{inits} program.}
\label{fig:notterm}
\end{figure}
Here the outer loop cuts away the last cell until the initial list consists only of one cell whereas the inner loop is used to iterate through the list.
It is easy to see that the main function terminates if the argument is an acyclic list.
Since variables $ys$ and $cur$ share during iteration, the proposed transformation introduces a fresh variable for the \ttt{next} field of the initial argument $ys$ when performing the $\jputfield$ instruction.
Termination of the resulting rewrite system can not be shown any more.

However \emph{non-termination preservation} follows as an easy corollary of Theorem~\ref{t:2}.
\begin{corollary}
\label{c:2}
The computation graph method, that is the transformation from
a given JBC program $\Program$ to a cTRS $\RS$ is non-termination preserving.  
\end{corollary}
\begin{proof}
Suppose there exists an infinite run in $\Program$, but
$\RS$ is terminating.
Let $\start$ be some initial state $\start$ of $\Program$. By Theorem~\ref{t:2}
there exists a state $t$ such that $\JVMexec{\start}{t}$ and
$\funsym{\astart}(\tst(\start')) \rstrew \funsym{\at}(\tst(t'))$, where
$\start \in \gamma(\astart)$, $\start' = \beta(\start)$, $t \in \gamma(\at)$, and $t' = \beta(t)$.
Furthermore, as $\RS$ is terminating 
we can assume $\funsym{\at}(\tst(t'))$ is in normalform.
However, as $t'$ is non-terminating, there exists a successor, 
thus Lemma~\ref{l:10} implies that $\funsym{\at}(\tst(t'))$ cannot
be in normalform. Contradiction.
\end{proof}

%% file: implementation.tex
\section{Implementation}
\label{Implementation}

A prototype, termed \jat, of the proposed method has been implemented in the Haskell programming language.
We use~\cite{SS05,RS06,GZ13} to provide acyclicity and reachability facts.

\begin{example}
  Figure~\ref{fig:flatten} depicts a slightly modified version of the motivating example from~\cite{OBEG10}. 
  The program $\ttt{flatten}$ collects all integers from a list of trees storing integers.
  The complexity tool $\tct$ is able to show that the rewrite system resulting from our proposed transformation has linear runtime complexity.
\end{example}
\begin{figure}[ht]
\centering
\begin{minipage}[t]{0.30\textwidth}
\begin{lstlisting}[firstnumber=1]       
class IntList{
  IntList next;
  int value;
}

class Tree{
  Tree left;
  Tree right;
  int value;
}

class TreeList{
  TreeList next;
  Tree value;
}
\end{lstlisting}
\end{minipage}
\begin{minipage}[t]{0.49\textwidth}
\begin{lstlisting}
class Flatten {
  IntList flatten(TreeList list)
  TreeList cur = list;
  IntList result = null;
  while (cur != null){
    Tree tree = cur.value;
    if (tree != null) {
      IntList oldIntList = result;
      result = new IntList();
      result.value = tree.value;
      result.next = oldIntList;
      TreeList oldCur = cur;
      cur = new TreeList();
      cur.next = oldCur;
      cur.value = tree.left;
      oldCur.value = tree.right;
    } else {
      cur = cur.next;
    }
  }
  return result;
}
\end{lstlisting}
\end{minipage}
\caption{The \ttt{flatten} program.}
\label{fig:flatten}
\end{figure}

Currently $\tct$ only provides limited support for cTRSs.
A meaningful experimental evaluation will be provided in the future.

%% file: conclusion.tex
\section{Conclusion and Future Work}
\label{Conclusion}

In this paper we define a representation  
of  JBC  executions  as \emph{computation graphs} 
from which we obtain a representation of JBC 
executions as \emph{constrained rewrite systems}. We precise the 
\emph{widening} of abstract states so that the representation
of JBC executions is provably finite. Furthermore, we
show that the resulting transformation is complexity preserving. 

As emphasised above our approach does not directly give rise to an automatable
complexity-preserving transformation, but for that requires 
an extension by annotation or a dedicated shape analysis~\cite{NS12}. 
However our main result applies to \emph{any} computable approximation of the transformation
and in particular it shows complexity preservation 
of the transformation proposed by Otto et al.~\cite{OBEG10}. 
Moreover, it allows for an easy incorporation of the existing
wealth of results on shape analysis present in the literature and
thus improves upon the modularity of the proposed transformational
approach. 

%
Future work will be dedicated towards new methods for
complexity analysis of cTRSs.

%% file: appendix.tex
\section{Semantics of Jinja Bytecode Instructions}

\label{a:semantics}
\newlength{\mlength}
\setlength{\mlength}{.2em}

{%
\small
\begin{align*}
    \mraisebox{\jload~\num} \quad & \infer{\jvmstate{}{\heap}{\jvmframe{\loc(n) \cons \opstk}{\loc}{\cname}{\mname}{\PC+1} \cons \myframes}}{\jvmstate{}{\heap}{\jvmframez \cons \myframes}}
   \\[\mlength]
   \mraisebox{\jstore~\num} \quad & \infer{\jvmstate{}{\heap}{\jvmframe{\opstk}{\loc\{n \mapsto \val\}}{\cname}{\mname}{\PC+1}\cons \myframes}}{\jvmstate{}{\heap}{\jvmframe{\val \cons \opstk}{\loc}{\cname}{\mname}{\PC} \cons \myframes}}
   \\[\mlength]
   \mraisebox{\jpush~\val} \quad & \infer{\jvmstate{}{\heap}{\jvmframe{\val \cons \opstk}{\loc}{\cname}{\mname}{\PC+1}\cons \myframes}}{\jvmstate{}{\heap}{\jvmframez\cons \myframes}}
   \\[\mlength]
   \mraisebox{\jpop} \quad & \infer{\jvmstate{}{\heap}{\jvmframe{\opstk}{\loc}{\cname}{\mname}{\PC+1}\cons \myframes}}{\jvmstate{}{\heap}{\jvmframe{\val \cons \opstk}{\loc}{\cname}{\mname}{\PC}\cons \myframes}}
\end{align*}
}%
We use $\ttt{BOp}$ together with $\otimes = \{+, - , \vee, \wedge, \geqslant, ==, \neq\}$ to define instructions $\jiadd$, $\jisub$, $\jor$, $\jand$, $\jcmpgeq$, $\jcmpeq$ and $\jcmpneq$.
{%
  \small%
\begin{align*}%
  \mraisebox{\ttt{BOp}} \quad & \infer{\jvmstate{}{\heap}{\jvmframe{v_2 \otimes v_1 \cons \opstk}{\loc}{\cname}{\mname}{\PC+1}\cons \myframes}}{\jvmstate{}{\heap}{\jvmframe{v_1 \cons v_2 \cons \opstk}{\loc}{\cname}{\mname}{\PC}\cons \myframes}}\\[\mlength]
   \mraisebox{\jnot} \quad & \infer{\jvmstate{}{\heap}{\jvmframe{\neg b \cons \opstk}{\loc}{\cname}{\mname}{\PC+1}\cons \myframes}}{\jvmstate{}{\heap}{\jvmframe{ b \cons \opstk}{\loc}{\cname}{\mname}{\PC}\cons \myframes}}
  \\[\mlength]
   \mraisebox{\jiffalse~i} \quad & \infer{\jvmstate{}{\heap}{\jvmframe{\opstk}{\loc}{\cname}{\mname}{\PC+i}\cons \myframes}}{\jvmstate{}{\heap}{\jvmframe{\false \cons \opstk}{\loc}{\cname}{\mname}{\PC}\cons \myframes}}
   \\[\mlength]
   & \infer{\jvmstate{}{\heap}{\jvmframe{\opstk}{\loc}{\cname}{\mname}{\PC+1}\cons \myframes}}{\jvmstate{}{\heap}{\jvmframe{\true \cons \opstk}{\loc}{\cname}{\mname}{\PC}\cons \myframes}}
   \\[\mlength]
    \mraisebox{\jgoto~i} \quad & \infer{\jvmstate{}{\heap}{\jvmframe{\opstk}{\loc}{\cname}{\mname}{\PC+i}\cons \myframes}}{\jvmstate{}{\heap}{\jvmframe{\opstk}{\loc}{\cname}{\mname}{\PC}\cons \myframes}}
\end{align*}%
}%
$\jnew~\cname'$ creates a new instance $obj$ of class $cn'$. 
The fields of $obj$ are instantiated with the default values, ie., $0$ for $\tyint$, $\m{false}$ for $\tybool$ and $\mynull$ otherwise.
Instance $obj$ is mapped to by a fresh address $a$ in $heap$.
$\jgetfield~fn~cn'$ access field $(cn', fn)$ of $\ftof(\heap(a))$.
$\jputfield~fn~cn'$ updates field $(cn', fn)$ in $(cn'', ftable) = \heap(a)$ with value $v$.
$\jcheckcast~cn'$ fails if $cn' \subclass cn$ does not hold. $\jgetfield$ and $\jputfield$ fail if $a$ is $\mynull$.
{%
  \small
  \begin{align*}
   \mraisebox{\jnew~\cname'} \quad & \infer{\jvmstate{}{\heap\{\addr \mapsto obj\}}{\jvmframe{\addr \cons \opstk}{\loc}{\cname}{\mname}{\PC + 1} \cons \myframes}}{\jvmstate{}{\heap}{\jvmframez \cons \myframes}}
   \\[\mlength]
   \mraisebox{\jgetfield~\fname~\cname'} \quad & \infer{\jvmstate{\varnothing}{\heap}{\jvmframe{\ftable(\cname', \fname) \cons \opstk}{\loc}{\cname}{\mname}{\PC + 1} \cons \myframes}}{\jvmstate{\varnothing}{\heap}{\jvmframe{\addr \cons \opstk}{\loc}{\cname}{\mname}{\PC} \cons \myframes}}
   \\[\mlength]
   \mraisebox{\jputfield~\fname~\cname'} \quad & \infer{\jvmstate{\varnothing}{\heap\{\addr \mapsto (\cname'', \ftable')\}}{\jvmframe{\opstk}{\loc}{\cname}{\mname}{\PC + 1} \cons \myframes}}{\jvmstate{\varnothing}{\heap}{\jvmframe{\val \cons \addr \cons \opstk}{\loc}{\cname}{\mname}{\PC} \cons \myframes}}
   \\[\mlength]
   \mraisebox{\jcheckcast~\cname'} \quad & \infer{\jvmstate{\varnothing}{\heap}{\jvmframe{cn \cons \opstk}{\loc}{\cname}{\mname}{\PC + 1} \cons \myframes}}{\jvmstate{\varnothing}{\heap}{\jvmframe{cn \cons \opstk}{\loc}{\cname}{\mname}{\PC} \cons \myframes}}
  \end{align*}
}%
$\jinvoke~mn'~n$ inspects the type of $\heap(a)$, and performs a bottom-up search (with respect to the subclass hierarchy) for the first method declaration $mn'$.
The new frame is $\myframe' = (\epsilon, \loc, cn', mn',0)$, where $loc$ consists of the $this$ reference (address $a$), parameters $p_0\cons \dots \cons p_{n-1}$ and $mxl$ registers instantiated with $\unit$ ($mxl$ is defined in the method declaration), and $cn'$ denotes the class where $mn'$ is declared.
The program terminates if $\jreturn$ is executed and $frms$ consists of a single frame.
Otherwise, the top frame is dropped and the next frame updated; $frm'$ drops the parameters and the reference and pushes the return value $v$ onto the stack.
{%
  \small
  \begin{align*}
   \mraisebox{\jinvoke~\mname'~n} \quad & \infer{\jvmstate{\varnothing}{\heap}{\myframe' \cons (p_{n-1} \cons \dots \cons p_{0} \cons a \cons \opstk,\loc,\cname,\mname,\PC) \cons \myframes}}{\jvmstate{\varnothing}{\heap}{(p_{n-1} \cons \dots \cons p_{0} \cons a \cons \opstk,\loc,\cname,\mname,\PC) \cons \myframes}}
   \\[\mlength]
   \mraisebox{\jreturn} \quad & \infer{\jvmstate{\varnothing}{\heap}{[]}}{\jvmstate{\varnothing}{\heap}{[\myframe]}}
   \qquad \infer{\jvmstate{\varnothing}{\heap}{\myframe' \cons \myframes}}{\jvmstate{\varnothing}{\heap}{\jvmframe{\val \cons \opstk}{\loc}{\cname}{\mname}{\PC} \cons \myframe \cons \myframes}}
  \end{align*}
}